\documentclass{aa}
\usepackage{graphicx}	% Including figure files
\usepackage{txfonts}
\usepackage{amsmath}	% Advanced maths commands
\usepackage{amssymb}	% Extra maths symbols
\usepackage{multicol}        % Multi-column entries in tables
\usepackage{bm}		% Bold maths symbols, including upright Greek
\usepackage{pdflscape}	% Landscape pages
\usepackage{color} % colored text
\usepackage{hyperref}
\usepackage{placeins}
\usepackage{booktabs}
\usepackage{multirow}
\usepackage{csquotes}
\usepackage[
  locale=US,
  separate-uncertainty=true,
  per-mode=symbol-or-fraction,
  range-phrase=-,
]{siunitx}
\usepackage{multirow}

\newcommand\uv[1]{$(u, v)$ #1}

\DeclareSIUnit{\arbitraryunit}{a.u.}
\DeclareSIUnit{\gigabyte}{GB}

\makeatletter
\renewcommand*\aa@pageof{, page \thepage{} of \pageref*{LastPage}}
\makeatother
\begin{document}
\renewcommand{\figureautorefname}{Fig.}
\renewcommand{\tableautorefname}{Tab.}
\renewcommand{\sectionautorefname}{Sect.}
\renewcommand{\equationautorefname}{Eq.}

   \title{Deep Learning-based Imaging in Radio Interferometry}
   
   \author{
    K. Schmidt\inst{1}\thanks{Contact e-mail: \href{mailto:kevin3.schmidt@tu-dortmund.de}{kevin3.schmidt@tu-dortmund.de}},
    F. Geyer\inst{1}\thanks{Contact e-mail: \href{mailto:felix.geyer@tu-dortmund.de}{felix.geyer@tu-dortmund.de}},
    S. Fröse\inst{1},
    P.-S. Blomenkamp\inst{1},
    M. Brüggen\inst{2},
    F. de Gasperin\inst{2,3},
    D. Elsässer\inst{1}
    \and\\
    W. Rhode\inst{1}
    }

   \institute{Lehrstuhl für Experimentelle Physik V, TU Dortmund University,
              Otto-Hahn-Straße 4a, 44227 Dortmund, Germany
         \and
             Hamburger Sternwarte, University of Hamburg,
             Gojenbergsweg 112, 21029 Hamburg, Germany
\and INAF - Istituto di Radioastronomia, via P. Gobetti 101, 40129 Bologna, Italy}
   \date{Received ; accepted }

% \abstract{}{}{}{}{} 
% 5 {} token are mandatory
 
  \abstract
   {The sparse layouts of radio interferometers result in an incomplete sampling of the sky in Fourier space which leads to artifacts in the reconstructed images.
   Cleaning these systematic effects is essential for the scientific use of radiointerferometric images.}
   {Established reconstruction methods are often time-consuming, require expert-knowledge, and suffer from a lack of reproducibility.
   We have developed a prototype Deep Learning-based method that generates reproducible images in an expedient fashion.}
   {To this end, we take advantage of the efficiency of Convolutional Neural Networks to reconstruct image data from incomplete information in Fourier space. The Neural Network architecture is inspired by super-resolution models that utilize residual blocks.
   Using simulated data of radio galaxies that are composed of Gaussian components we train Deep Learning models whose reconstruction capability is quantified using various measures.}
   {
   %We use the predicted Fourier data to calculate the sources' brightness distributions by applying the inverse Fourier transformation leading to small deviations between reconstructed and true source images.
   The reconstruction performance is evaluated on clean and noisy input data by comparing the resulting predictions with the true source images. We find that source angles and sizes are well reproduced, while the recovered fluxes show substantial scatter, albeit not worse than existing methods without fine-tuning. Finally, we propose more advanced approaches using Deep Learning that include uncertainty estimates and a concept to analyze larger images.
%   The comparison of the source areas results in a mean ratio of \num{0.98\pm0.13} and \num{0.96\pm0.17}, the mean deviation of the jet angles calculates to \SI{0.03\pm1.52}{\degree} and \SI{-0.11\pm2.40}{\degree}, and the flux densities for the core components have a mean deviation of \SI{-0.01\pm0.12}{\percent} and \SI{-0.02\pm0.13}{\percent} for clean and noise input data, respectively.

   %All simulation and analysis steps can be applied using the \texttt{radionets} framework, which is available for the astronomy community as an open-source package.
   }
   {}

   \keywords{Galaxies: active -- Radio continuum: galaxies -- Methods: data analysis -- Techniques: image processing -- Techniques: interferometric
               }

   \titlerunning{Deep Learning-based Imaging in Radio Interferometry}
   \authorrunning{Schmidt et al.}
   \maketitle

\section{Introduction}
\label{sec:introduction}

With radio interferometry, it is possible to obtain images of radio sources with angular resolutions of up to milli-arcseconds \citep{milli-arc}.  Achieving such high angular resolutions was made possible by the advent of Very Long Baseline Interferometry (VLBI) \citep{vlbi_broten}, which exploits large distances between two telescopes. Most current radio telescopes are radio interferometers, such as the VLA, LOFAR or MeerKAT. Furthermore, data from radio interferometers play a decisive role in multiwavelength studies \citep{mwl_lena, mwl_gw}.

Radio interferometers record information about the sky in Fourier space, also called visibility space. The relation between the measurement and the specific brightness distribution of the source is described by the van Cittert-Zernike theorem \citep{Cittert-Zernike, ri_cittert}, which states that the two-point correlation function of the electric field measured by two antennas of a radio interferometer is the Fourier-transformed intensity distribution of the source. As the number of antennas in a radio interferometer array is limited, the sampled Fourier space always remains incomplete. By applying the inverse Fourier transformation to the data, artifacts dominate the reconstructed image. For this reason, data first needs to be "cleaned", by the astronomer to use them for scientific analyses \citep{clean, ri_clean}.

Established cleaning software such as DIFMAP \citep{difmap} or CASA \citep{casa} partly require human intervention during the analysis, for example marking the area of the sky where emission is expected ("masking"). This process is iterative, slow, and generates non-reproducible results because the reconstructed images depend on the user's experience. More recent approaches such as WSCLEAN \citep{wsclean} perform the imaging faster and with a higher degree of automatization. Nevertheless, the analysis is still time-comsuming since several parameters have to be adjusted in iterative cleaning runs in order to determine the best-suited parameter set for the given data quality. With increasing data rates of modern radio interferometers such as LOFAR \citep{lofar} and the SKA \citep{ska}, fast solutions are necessary to analyze observations on reasonable timescales. 

With increasing computing power, Deep Learning-based analysis strategies become more widely used in astronomy and astroparticle physics. Neural networks are commonly used because they are successful in other domains, their application is reasonably fast, and they generate reproducible results. First attempts of application to data from radio interferometers have already been tried in \cite{rim1} and \cite{rim2}. However, they often remain black boxes and their output is not always easy to interpret.

In this work, we use Convolutional neural networks (CNNs) because these networks have proven to be efficient tools for image tasks.
More precisely, we propose a CNN built from elements used in the context of super-resolution applications. Super-resolution networks have the purpose of converting low-resolution images into high-resolution versions by upsampling and reconstructing the fine-scale structures \citep{superr}.
These networks use the available information and enhance it to perform their assignments.
This is possible because the convolution kernels use values of the neighboring pixels to determine a value for the pixel to be estimated.
This way, the missing information can be reconstructed and corrupted pixel values corrected.
When analyzing data from radio interferometers, we face a similar problem since the visibility space is incompletely sampled. This procedure is comparable to the reconstruction of additional pixels in the upsampling process.
Thus, we aim to solve the reconstruction problem with an architecture inspired by \cite{superres}. 

 We show that our approach offers a fast, reproducible way to generate clean radio images by evaluating the reconstruction quality using simulated data. 
We developed a \texttt{radionets} framework \citep{radionets} that is available for the astronomical community as an open-source package. We provide a helpful tool to speed up the imaging process in radio interferometry, which does not rely on user input during the training and the application process.

Finally, we propose more advanced strategies to improve the simulations and the cleaning of the data. We upgrade our simulations by including point sources in the images creating more complex data sets. Moreover, we present ideas to how to deal with noise and larger images. By adding uncertainty estimates for the reconstructed source images, one can attain more meaningful results. Moreover, source finders based on Deep Learning networks can help to speed up the imaging process in general.

\autoref{sec:simulations} motivates and describes the simulations that were used to create the training data. 
The architecture and parameters for the training of the neural networks are laid out in \autoref{sec:training}.
In the next section, the results are evaluated. In \autoref{sec:reconstruction} we compare a network trained on clean and a network trained on noisy input data to the results of the established cleaning method \texttt{wsclean}.
\autoref{sec:further_approaches} lists additional approaches and ideas for further analysis.
In \autoref{sec:conclusions} we present our conclusions.
\begin{figure}
    \centering
    \includegraphics[width=\hsize]{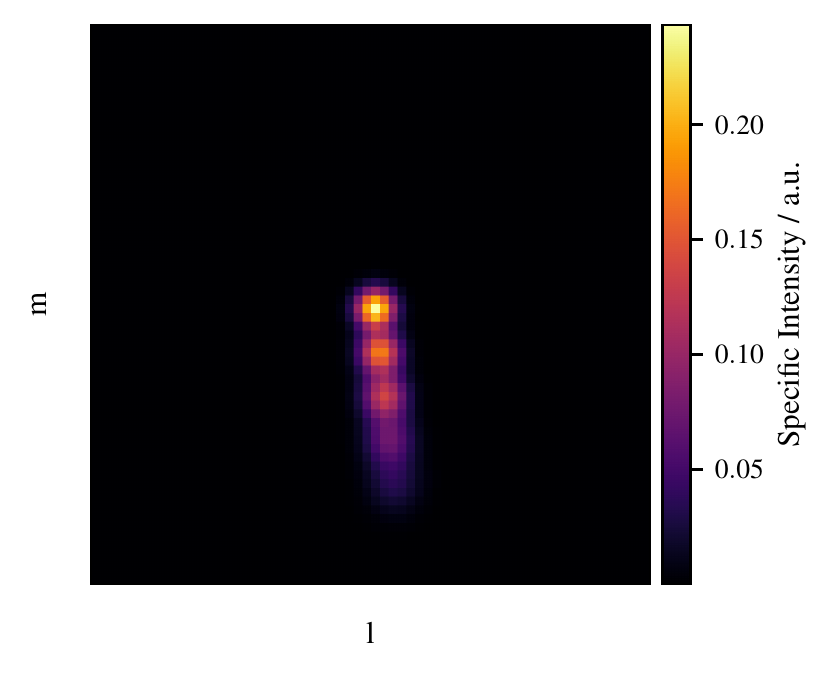}
    \caption{Image of a radio galaxy from the simulated data set in image space. The source is composed of two-dimensional Gaussian distributions, which are blurred with a Gaussian kernel.}
    \label{fig:source}
\end{figure}

\begin{figure}
    \centering
    \includegraphics[width=\hsize]{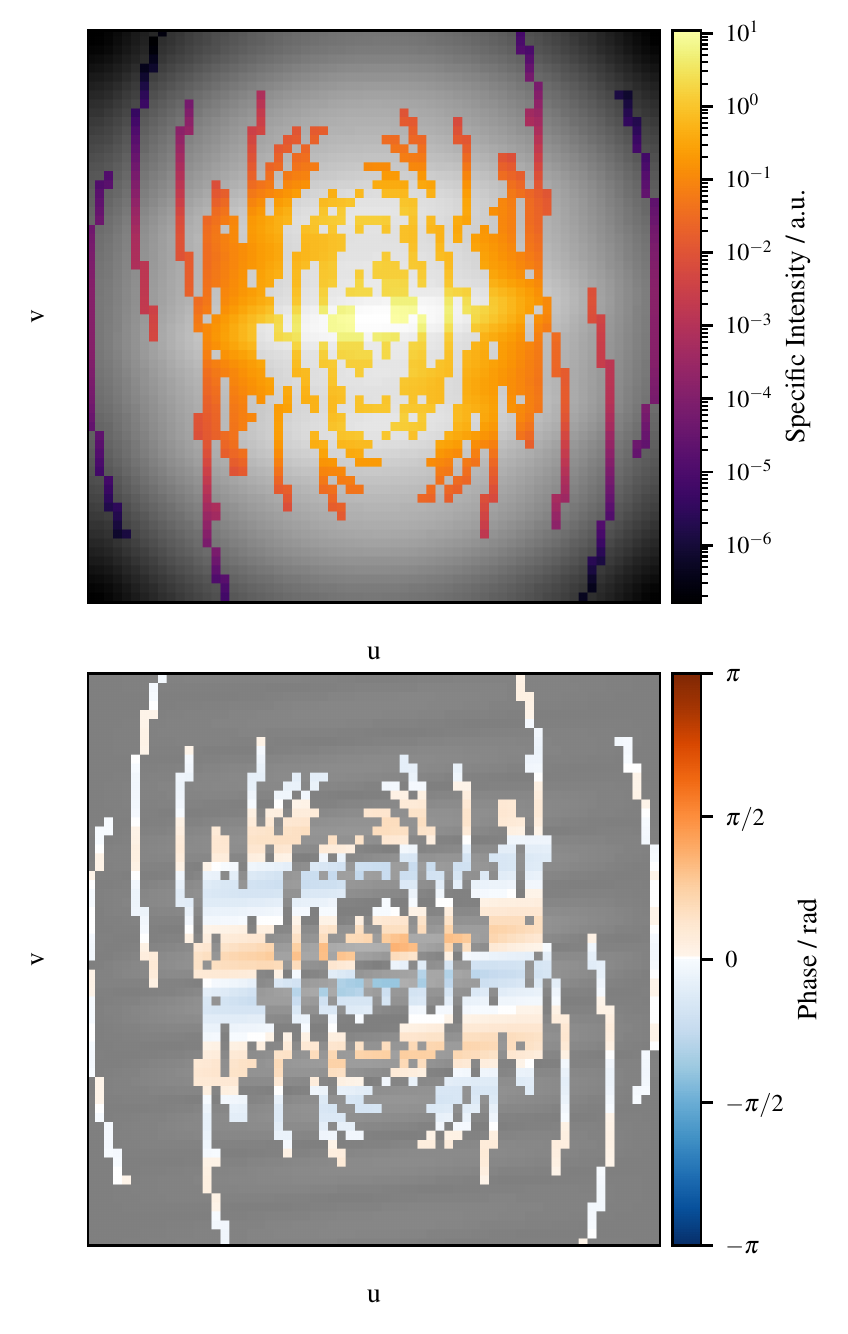}
    \caption{Amplitude (top) and phase (bottom) of the simulated radio galaxy. In the first step, the representation of the radio galaxy in Fourier space was calculated by Fourier transform. Then the complex values were transformed to amplitude and phase utilizing Euler's formula. To mimic an observation with a radio interferometer, the images were sampled using the \uv coverage of a simulated observation with the VLBA. The remaining information is displayed in color, all dropped information is grayed out.}
    \label{fig:freqs_samp}
\end{figure}

\section{Simulations}
\label{sec:simulations}

In order to train Deep Learning models, we create synthetic data with Monte-Carlo simulations of known ground truths \citep{hastie}.
Since the creation of realistic simulations is a time consuming and elaborate process, we start with simple simulations of radio galaxies. The goal is to extend and improve the simulations within the ongoing process of this project, see \autoref{subsec:improved-simulations}.

\begin{figure}
    \centering
    \includegraphics[width=\hsize]{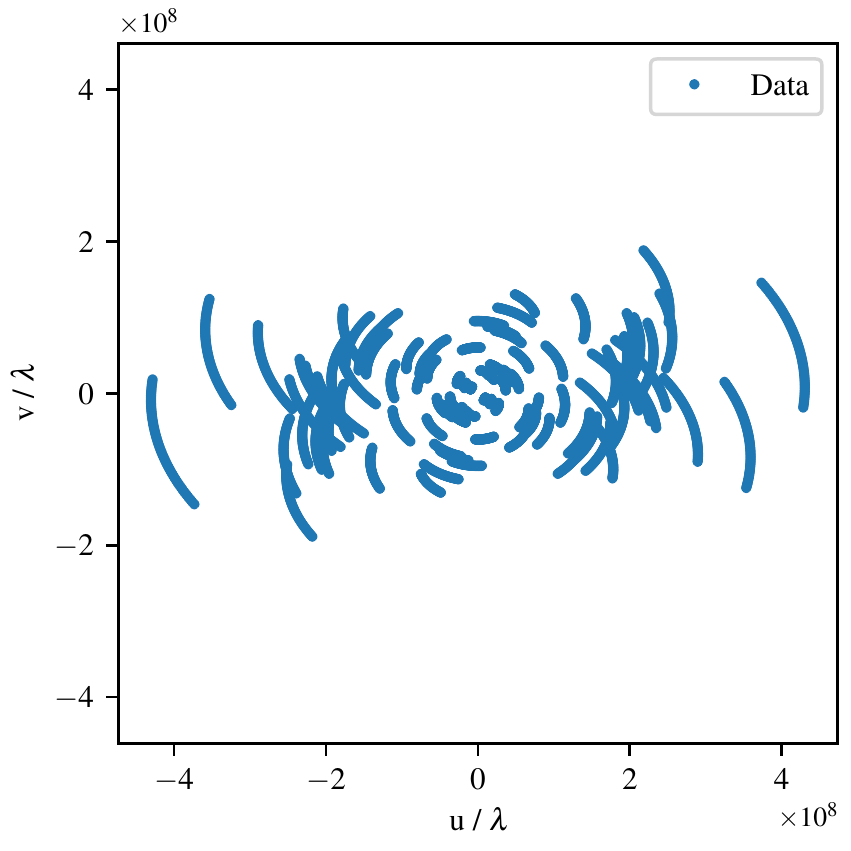}
    \caption{Exemplary \uv coverage for a simulated observation with 50 time steps.}
    \label{fig:projection}
\end{figure}

\subsection{Extended Radio Galaxies}

One of the targets of VLBI observations are radio galaxies, which are a subclass of active galactic nuclei.
Radio galaxies typically consist of a bright core and two jets that emerge from the central black hole region \citep{radiojets}. Because of relativistic boosting, these jets can also appear one-sided, depending on the angle of observation  \citep{urry}.
%In these cases, the counter jet is not visible for the observer because of the relativistic boosting of the particles.

According to the model of Blandford and Königl \citep{bland_koenigl, koenigl}, active galactic nuclei host narrow conical jets fueled by continuous plasma inflows.
%Shock fronts form in these jets, which differ from the extended structure by their increased luminosity \citep{spada}.
Using these so-called jet components for the source representation is a common simplification in the analysis of VLBI data, see e.g. \cite{gaussian_1, gaussian_2}.
Hence, we simulate radio galaxies by two-dimensional Gaussian distributions.
\autoref{fig:source} shows an example of such a source where the brightest component, called the core component, is always placed in the center of the image.
For variations within the data set, the following source parameters are randomly drawn:
First, the number of one-sided jet components is chosen between 3 and 6.
This ensures that the jet can reach the edge of the image, but not beyond.
An angle for the jet orientation is randomly drawn from a range between $\SI{0}{\degree}$ and $\SI{360}{\degree}$.
A straight line starting from the core component and moving towards the edge of the image is used to place the centers of the jet components.
The distance between two jet components is fixed to a value of 5 pixels.
Furthermore, the components are initialized with random amplitudes.
The core components' peak amplitudes range between $10^{-3}$ and $10^{1}$ and are drawn randomly from a uniform distribution. 
Jet components' peak amplitudes are logarithmically reduced by calculating ${A_\text{core}} / {\text{exp}(n_\text{comp})}$.
Additionally, we use randomized standard deviations with increasing $\sigma_\text{x}$ and $\sigma_\text{y}$ for components further outside the jet.
In the case of a two-sided jet, the selected jet components are mirrored at the core component, leading to a total number of components between 4 and 13.

\subsection{Observations with Radio Interferometers}
\label{sec:obs}

Radio interferometers 
measure complex values in \uv{space}, which offers the possibility to reconstruct the brightness distribution of the source using the inverse Fourier transform \citep{ri_intro}.
Using Euler's formula \citep{euler} to transform complex data into amplitudes and phases limits the ranges in the data, which helps in training Deep Learning models. For illustration,
\autoref{fig:freqs_samp} shows amplitudes and phases for an example of a simulated radio galaxy.

Limited by the antenna configuration, the entire \uv plane is never completely sampled.
In the case of the VLBA, the ten antennas lead to 90 baselines \citep{vlba}.
Each baseline provides one complex value per snapshot.
Utilizing the Earth's rotation helps to fill the \uv plane as densely as possible \citep{rotation}.
Before correlating the data, one has to compensate for the arrival time difference between individual antennas caused by the geometric layout of the interferometer.
This is done by defining a reference point on a plane perpendicular to the source direction.
All arrival times of the signals are adjusted to form a two-dimensional plane containing all antennas. 
Owing to the Earth's rotation, baselines change with respect to the source direction, which in turn provides additional data points in \uv space.

\autoref{fig:projection} shows the \uv coverage of a simulated single channel observation in which
large areas of the \uv space remain unsampled.
We create such sampling masks for the complete \uv plane to simulate the incomplete coverage by radio interferometer. Thus, data sets can be created with a variety of \uv coverages.
It is straightforward to change different observation parameters such as the starting hour angle of the source and the length of the observation.
The maximal baselines correspond to resolutions of approximately $2\,\text{pixels}$ and the smallest baselines allow to resolve scales of the order of approximately $8\,\text{pixels}$.
Note that the scales given here represent values in units of the image size of $64\,\text{pixels}$ and do not refer to absolute angles.
In the simulation in the \uv space, instrumental effects of the radio interferometer are ignored. The resulting \uv data are comparable to already gridded visibility data $V_\text{noiseless}$. In the following, we will refer to frequels as gridded data points in Fourier space, in the same way as pixels are used to describe the source distributions in image space. Simulations of un-gridded data that include antenna characteristics can be performed using the RIME formalism \citep{rime}, which is discussed in \autoref{subsec:improved-simulations}.

\begin{figure*}
    \centering
    \includegraphics[width=0.495\hsize]{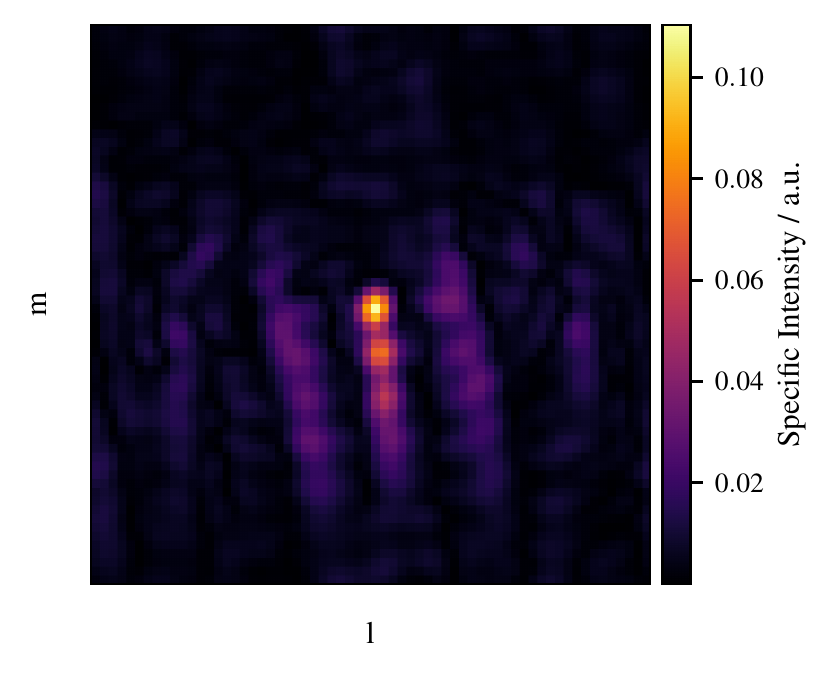}
    \includegraphics[width=0.495\hsize]{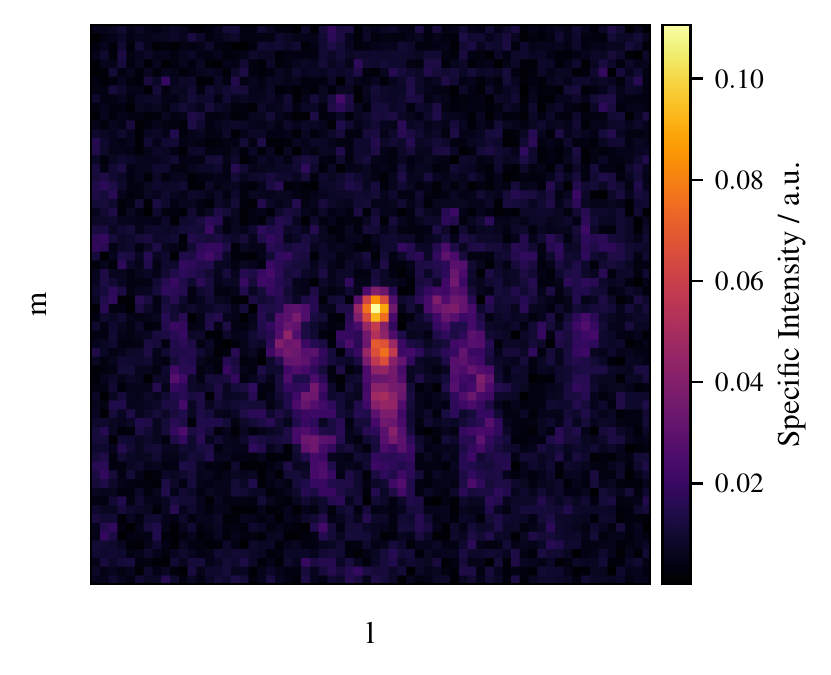}
    \caption{Result of the inverse Fourier transformation applied to the sampled noiseless visibilities (left) and to the noisy visibilities with additional white noise (right), called \enquote{dirty images}. The appearance of background artifacts is visiblein both cases. Compared to the true brightness distribution shown in \autoref{fig:source} the brightness is underestimated. These reconstruction errors are caused by incomplete data in Fourier space.}
    \label{fig:recons_source}
\end{figure*}

\autoref{fig:freqs_samp} shows the amplitude and phase distribution for the example source, where all information lost in the sampling process are grayed out and set to zero for the next data processing steps.
This example illustrates the underlying problem of reconstructing missing radio interferometric data.
If these incomplete \uv spaces are used to reconstruct the source's brightness distribution by inverse Fourier transformation, so-called "dirty images" dominated by artifacts are generated illustrated in \autoref{fig:recons_source}.
The goal of our Convolutional Neural Network is to clean up these dirty images or to reconstruct the incomplete data before applying the inverse Fourier transformation.
The colored parts of the \uv space shown in \autoref{fig:freqs_samp} are the input data for the neural networks, and the complete Fourier planes, as shown in gray, represent the target images.

\begin{figure}
    \centering
    \includegraphics[width=\hsize]{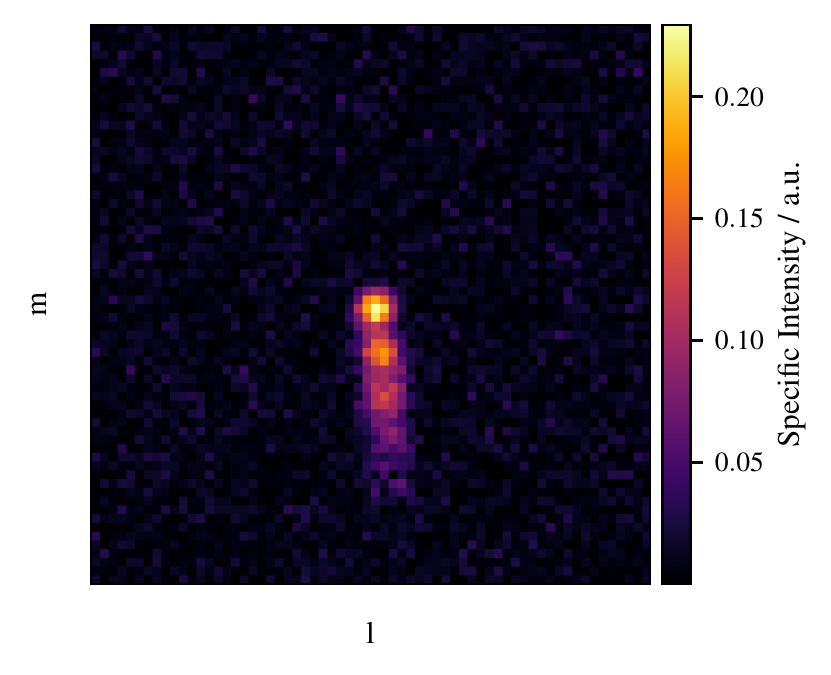}
    \caption{Image of a radio galaxy from the simulated data set in image space with added noise corruption $I_\text{noisy}$. The source is composed of two-dimensional Gaussian distributions, which are blurred with a Gaussian kernel.}
    \label{fig:source_noise}
\end{figure}

\subsection{Uncorrelated Noise}
\label{sec:noise}

%In addition to signals, real observations  include noise.
In order to mimic noise, we smear the pixel values of the simulated radio galaxies in image space by adding offsets for every pixel.
These offsets are randomly drawn numbers from a standard normal distribution:

\begin{align}
    g(x~|~\mu, \sigma) = \frac{1}{\sqrt{2\pi\sigma^2}} \exp\left(-\frac{(x - \mu)^2}{2\sigma^2}\right).
\end{align}
Before adding the random noise, the values get scaled by a factor of $5\,\%$ of the peak source flux resulting in noisy images given by
\begin{align}
    I_\text{noisy}(l, m) = I(l, m) + I_\text{max} \cdot \boldsymbol{g(x~|~\mu, \sigma)_{lm}}.
\end{align}
Here, $I(l, m)$ represents the image intensity at pixel $(l, m)$, $I_\text{max}$ stands for the peak intensity of the image, and $g(x~|~\mu, \sigma)_{lm}$ for a new randomly drawn number at pixel coordinate $(l, m)$ with $\mu=0$ and $\sigma=1$.
This procedure results in corrupted Fourier data $V_\text{noisy}$ when calculating the Fourier transform.
\autoref{fig:source_noise} shown an example of a simulated radio galaxy with added noise.

To increase the complexity of our data and to simulate possible measurement effects in \uv{space}, we add additional white noise to the visibilities. The white noise is produced by drawing random values from a Gaussian distribution with zero mean and a standard deviation of 1 resulting in corrupted visibilities

\begin{align}
    V_\text{noisy+white noise}(u, v) = V_\text{noisy}(u, v) + \boldsymbol{g(x~|~\mu, \sigma)_{uv}}.
\end{align}
Again, $\boldsymbol{g(x~|~\mu, \sigma)_{uv}}$ is a new random number for every frequel with $\mu=0$ and $\sigma=1$.
\autoref{fig:recons_source} illustrates the dirty image created from the noisy visibilities with additional white noise on the right. Additional artifacts caused by the noise corruption are visible.
In \autoref{sec:noisy_input}, $V_\text{noisy}$ and $V_\text{noisy\&white noise}$ are both used separately as input data for our neural network and the reconstruction results are compared.
This noise is uncorrelated while real observations can suffer from substantial correlated noise. In the future, we plan to consider also correlated noise as described in \autoref{subsec:improved-simulations}.

\section{Model Training}
\label{sec:training}

% Here we begin by describing the code architecture that deals with the issue of incomplete \uv spaces.
In this section, we will describe our neural network. Specifically, we will illustrate the architecture and its components, describe the input data and the data augmentation applied to it, explain our loss function and present the optimizer function, which is used for the minimization process. %The role of the neural network is to reconstruct the missing pixels in Fourier space such that the corresponding image is optimally reproduced.

\subsection{Architecture}

\begin{figure*}
    \centering
    \includegraphics[width=\hsize]{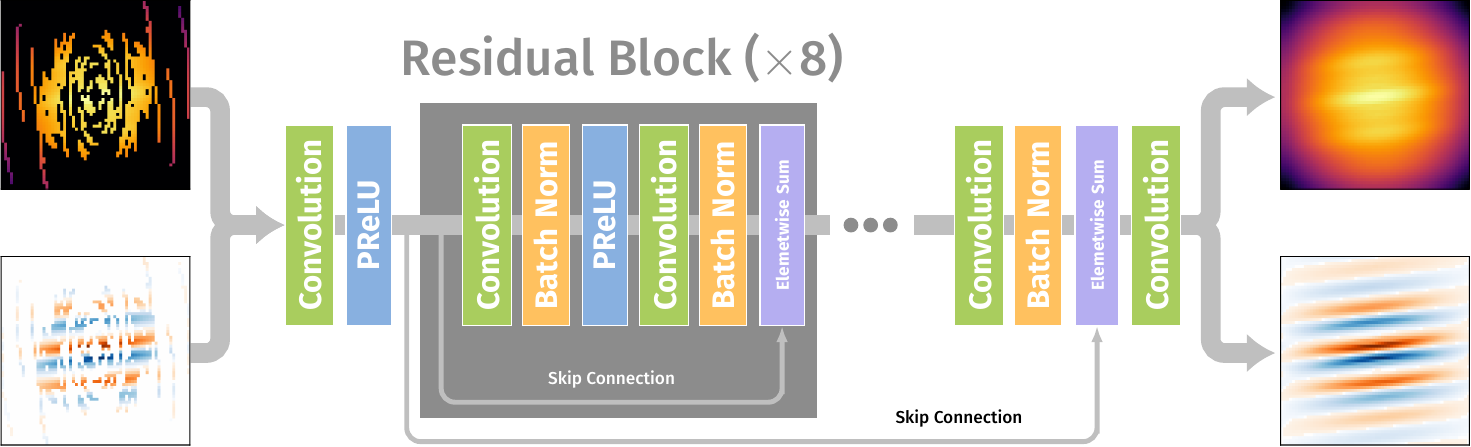}
    \caption{Schematic representation of the architecture employed in this paper. Sampled amplitude and phase distributions serve as input. After a pre-convolution followed by a PRelu activation \citep{prelu}, the main part of the architecture is composed of eight residual blocks \citep{residual-nets, residual-block}. One block consists of two convolution batch norm pairs with a PRelu activation in between. The input data is by-passed and added element-wise to the output for every residual block. The last residual block is followed by a convolution batch norm pair. Another skip connection is placed around the main part of the architecture. After a final convolution the reconstructed amplitude and phase distributions are the output of the architecture. Diagram reproduced from \citep{superres}.}
    \label{fig:resblock}
\end{figure*}

\begin{table*}[ht!]
    \centering
    \caption{Overview of the used architecture. The properties of the different stages are described. For each layer, the number of input and output channels is given. In the case of convolutional layers, the kernel size, stride, and padding are specified.}
    \begin{tabular}{llccccc
    }
    \toprule
    {stage} & {layer} & {channel in} & {channel out} & {kernel size} & {stride} & {padding} \\
    \midrule
    \multirow{2}{*}{Pre Block} & Convolution & 2 & 64 & $(9 \times 9)$ & 1 & 4 \\
    & PReLU & 64 & 64 & - & - & - \\
    \midrule
    \multirow{6}{*}{Residual Block ($\times$8)} & Convolution & 64 & 64 & $(3 \times 3)$ & 1 & 1 \\
    & Batch Norm & 64 & 64 & - & - & - \\
    & PReLU & 64 & 64 & - & - & - \\
    & Convolution & 64 & 64 & $(3 \times 3)$ & 1 & 1 \\
    & Batch Norm & 64 & 64 & - & - & - \\
    & Elementwise Sum & 64 & 64 & - & - & - \\
    \midrule
    \multirow{3}{*}{Post Block} & Convolution & 64 & 64 & $(3 \times 3)$ & 1 & 1 \\
    & Batch Norm & 64 & 64 & - & - & - \\
    & Elementwise Sum & 64 & 64 & - & - & - \\
    \midrule
    Final Block & Convolution & 64 & 2 & $(9 \times 9)$ & 1 & 4 \\
    \bottomrule
    \end{tabular}
    \label{tab:arch}
\end{table*}

The convolutional layers exploit the spatial correlation in images by passing a kernel with specified weights over the image. For most applications, the result is a down-sampled version of the input image, which contains some features extracted from the original image. In our case, we choose the parameters of our convolutional layers such that the image size does not change. Thus we use the available information from the sampled amplitude and phase maps, which serve as our input images, to reconstruct values for the missing information. This idea is sketched in \autoref{fig:resblock}. 

In super-resolution applications, high-resolution images are produced from low-resolution input images by up-sampling and reconstructing the fine-scale structures. Using a combination of convolutional layers, a pixel value is determined based on the values of the neighbouring pixels. 
Applying this concept to the \uv plane, we
build a network following \cite{superres} using the residual block layout investigated by \cite{residual-nets} and adapted by \cite{residual-block}.
In this setup, every residual block consists of five operations.
In the first step, the input data passes through a convolutional layer with $64$ input and $64$ output channels, a kernel size of $(3 \times 3)$ pixels, a stride of $1$, and a padding of $1$. Kernel size refers to the extent of the kernel which is used to scan the image. The stride is the number of pixels, which lie between the centers of two individual convolutions. Furthermore, padding is used to add pixels to the edges of the images, which are filled with zeros to keep the same size for the output image.
Note that with these settings the image size remains the same as the input image size.
The second and third operation consist of batch normalization and a parametric rectified linear unit (PReLu) \citep{prelu} as non-linearity.
Here, the negative part of the function is not constant, but its coefficient is learned by the network.
The block ends with another convolution with the same settings as described above and an additional batch norm layer.
In parallel, the input data is bypassed through a skip connection and added element-wise to the output of the residual block.
This changes the underlying mapping function $F$ which the neural network has to learn to map the input $x$ to the output $y$:
\begin{align}
    F(x) = y - x.
\end{align}
Here, the network must only learn to map the difference between input and output, meaning that each block predicts only a residual.
Predicting only the residual leads to a faster convergence of the network and makes it more robust to outliers.
\autoref{fig:resblock} illustrates the layout of the residual blocks, which is highlighted by the gray box, as well as the complete architecture. In the following, we describe the properties of the different stages.

\begin{figure*}
    \centering
    \includegraphics[width=\hsize]{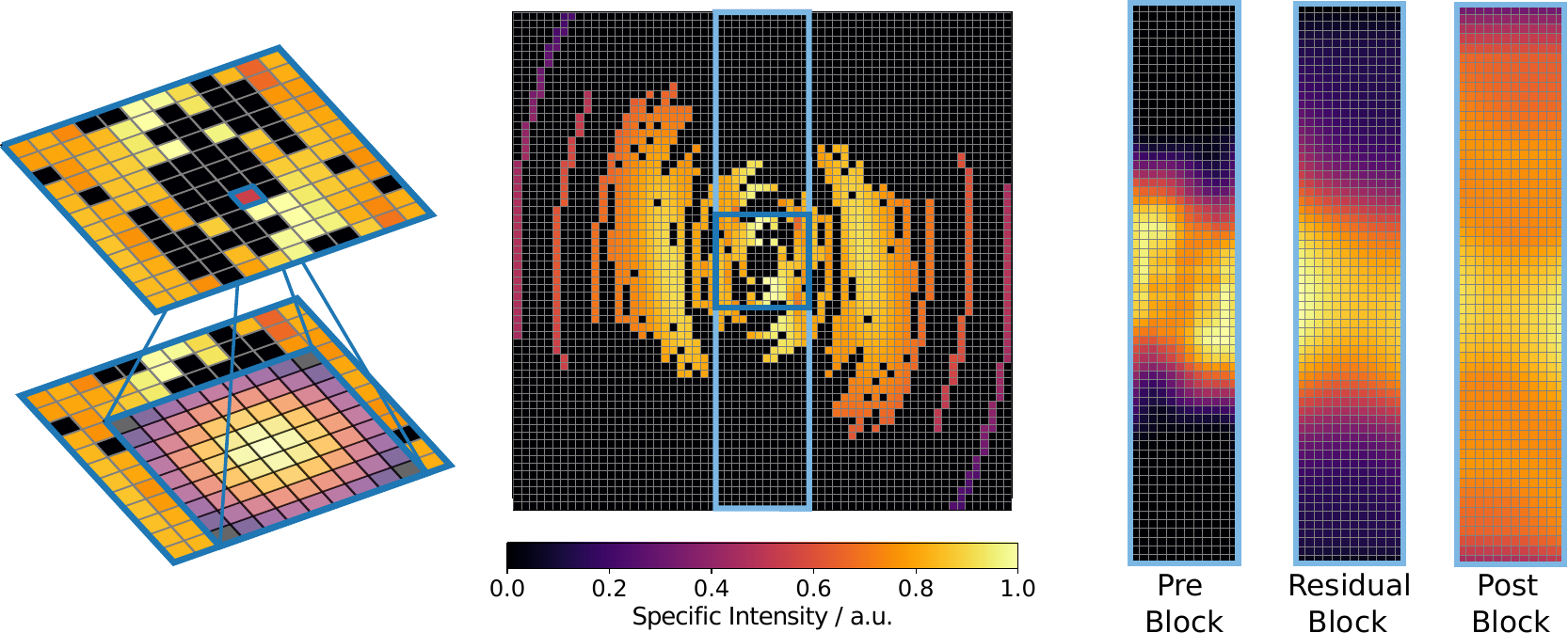}
    \caption{Schematic illustration how our neural network reconstructs incomplete amplitude and phase maps. Frequels with missing information have values of zero. Convolutions inside the architecture use information of neighboring frequels to calculate a new value for the center frequel. A convolution with a $(9 \times 9)$ kernel is shown on the left (dark-blue). An input amplitude map is shown in the middle. The part of the map used for the convolution example is marked with dark-blue. The part marked in light-blue is used to demonstrate the reconstruction progress after the different network stages. The reconstruction status after the Pre Block, Residual Block and Post Block is shown on the right. The amount of frequels without information decreases after every stage. The amplitude map is reconstructed up to the edges when it has passed through the Post Block. For simplicity, only the influence of the convolutions is taken into account. A re-scaling of the output of the different stages was performed for better visualization.}
    \label{fig:conv_filling}
\end{figure*}

The neural network starts with a pre-convolution taking two input channels and extending them to 64 output channels with a $(9 \times 9)$-kernel.
Using a padding of $4$ keeps the image size the same.
For this convolutional layer, we are using a group setting, which divides the parameters into two parts.
Thus, the output results in $32$ channels dedicated to the amplitude, and $32$ for the phase, both with individual filter weights.
The central part of the architecture consists of $8$ residual blocks. 
This part follows for an additional post-block with a $(3 \times 3)$ convolution and a batch norm layer.
The chosen settings keep the channel and image size the same.
An additional skip connection between the pre-convolution and the final convolution accelerates the convergence of the network since again only the residual has to be learned.
The final convolutional layer has the same settings as the pre-convolution.
This time, it takes the $64$ channels as input and making two channels out of it.
Again, $32$ channels are dedicated to the amplitude and $32$ channels to the phase by enabling the parameter grouping.
A summary of the architecture including the settings for the different layers can be found in \autoref{tab:arch}.

\subsection{Training}
For the network training, we use a data set that is simulated using the procedure described in \autoref{sec:simulations}.
For every simulated source in image space, we generate a different observation resulting in a new sampling mask for the corresponding \uv space.
This approach creates variation in the data set and serves as a form of data augmentation.
More importantly, the use of different sampling masks allows for a more realistic training process.
Furthermore, our experience shows that when using different sampling masks, the loss drops more than when using a static mask.
As stated in \autoref{sec:obs}, we transform the complex data into amplitude and phase maps using Euler's formula. In contrast to the real part, the amplitude distribution has only positive values.
Furthermore, the transformation of the imaginary part has a big impact on the training. The phase distribution is limited to a range from $-\pi$ to $\pi$, which is an enormous reduction of the parameter range to be learned, since the imaginary part distribution would span several orders of magnitude for the entire data set. This results in smoother training processes, faster convergence of the models, and improved fine-scale reconstructions. In general, the limitation of the parameter space has proven to be a great advantage when training neural networks.
The properties of the different data sets are as follows: 50\,000 training amplitude and phase maps, 10\,000 validation amplitude and phase maps, and 10\,000 test amplitude and phase maps.
In the input data, frequels with missing information are set to a value of 0 resulting in a range between $73\,\%$ and $82\,\%$ of all input frequels, which varies depending on the simulated observation.
During the training process we use data augmentation to prevent an overfitting of the network.
In the batch creation step, input and target maps are rotated by a random multiple of 90 degrees which further increases the number of individual training images.

Contrary to most other approaches, we train our network in Fourier space. The neural network is used to reconstruct missing data in the visibility space, such that a clean image can be generated using the Fourier transformation afterwards. The convolution filters use information from sampled frequels to calculate values for neighboring frequels with missing information, which are marked with zero values. This procedure is visualized in \autoref{fig:conv_filling}. The advantage of marking frequels with missing information with zeros is that these frequels do not distort the result of the convolutional layers. At the same time, zero pixels have a large offset to the target value of the simulated true amplitude and phase maps, which causes large losses. As a result, the network learns to reconstruct these pixels in a prioritized manner, as good reconstructions rapidly reduce the calculated loss. Step by step, the convolutional layers can use existing and newly filled information to calculate values for all frequels. In this way, it is possible to fill the complete amplitude and phase maps with continuous depth of the architecture. A direct advantage of this approach is that it is not necessary to switch between Fourier space and image space, as is the case between every iteration when using the CLEAN algorithm. In our case, no flux components are extracted from the dirty image to create a model of the radio source, but the information in Fourier space is used directly to reconstruct the missing information. Furthermore, it is not necessary to convolve the reconstructed images with a telescope beam, as we do not perform a point source extraction.
Another advantage of reconstructing in Fourier space is the ability to transfer knowledge gained on smaller amplitude and phase maps to be able to reconstruct larger ones with little additional effort, see \autoref{subsec:image-size}. This is possible because the field of view of the clean image is directly related to the spacing of the samples in Fourier space. 
%More information on the idea to transfer the learning on smaller images to larger ones is given in \autoref{subsec:image-size}.
Finally, working in Fourier space allows for an estimation of an uncertainty in the reconstructed values. %In addition to the reconstruction of missing values, the convolution kernels can be used simultaneously to estimate uncertainties for the respective frequels. 
This feature will be available in a future version of our network. The quantification of uncertainties also makes it possible to compute an uncertainty map for the clean image of the radio source. First tests have already been performed, see \autoref{subsec:uncertainty} for details.

\begin{figure}
    \centering
    \includegraphics[width=\hsize]{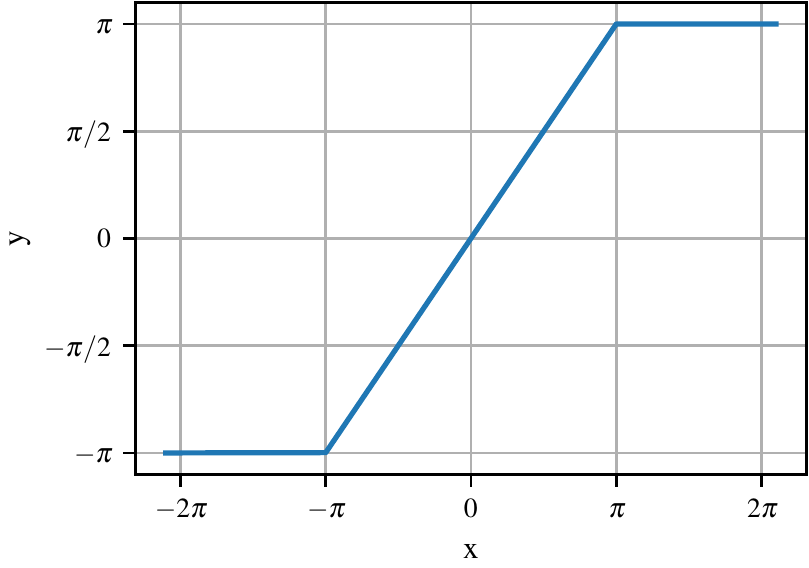}
    \caption{Illustration of the HardTanh$(x)$ function. Input data $x$, is mapped to a parameter range between $-\pi$ and $\pi$.}
    \label{fig:hardtanh}
\end{figure}

First, we train a network with noiseless input data $V_\text{noiseless}$ and then we investigate the influence of noise.
To this end, we train another network with noisy input data $V_\text{noisy}$, as described in \autoref{sec:noise}.
Both networks are trained for $300$ epochs with a batch size of $64$ and a learning rate of $2\cdot10^{-4}$.
This learning rate selection follows the evaluation of the result from the learning rate finder created by \cite{lrfind}.
For the loss function, we use an adapted L1 loss 
\begin{align}
    \mathrm{Loss} &= \mathrm{L1}\left(x_{\mathrm{amp}}, y_{\mathrm{amp}}\right) + \mathrm{L1}\left(\mathrm{HardTanh}(x_{\mathrm{phase}}), y_{\mathrm{phase}}\right),  \label{eq:l1} \\
    \mathrm{with} &~~\mathrm{L1}(x, y) = \left|~x - y~\right|,\\
    \mathrm{and} &~~\mathrm{HardTanh}(x) = \left\{%
    \begin{array}{ll}
        \pi, & \hbox{if x > $\pi$} \\
        -\pi, & \hbox{if x < $-\pi$} \\
        0, & \hbox{otherwise} \\
    \end{array}%
    \right.,
\end{align}
where $x$ is the predicted output from the network and $y$ as the true image.
The HardTanh$(x)$ function restricts the prediction for the phase to a parameter range between $-\pi$ and $\pi$, which is illustrated in \autoref{fig:hardtanh}. This avoids problems that can arise from the phase's periodic nature when reconstructing the values.
We chose the L1 loss over the MSE loss because our experiments showed that the reconstruction of fine details and small scales were less accurate when using the MSE loss.
In our case,  the slightly adapted L1 loss as shown in \autoref{eq:l1} leads to improvements in the reconstructed \uv spaces which in turn result in more detailed and cleaner reconstructed source images.

We use the ADAM optimizer \citep{adam} to update the weights during training, which outperforms stochastic gradient decent (SGD) \citep{sgd} in convergence time at the cost of more readily learned parameters.

Our training time is about 170 seconds for one epoch on computer specifications described in \autoref{tab:computer-specifications}.
The complete training of 300 epochs thus took just over 14 hours.
The application time of the trained neural network is much lower, on the order of milliseconds per image and will be further discussed in \autoref{sec:reconstruction}.

\begin{figure}
    \centering
    \includegraphics[width=\hsize]{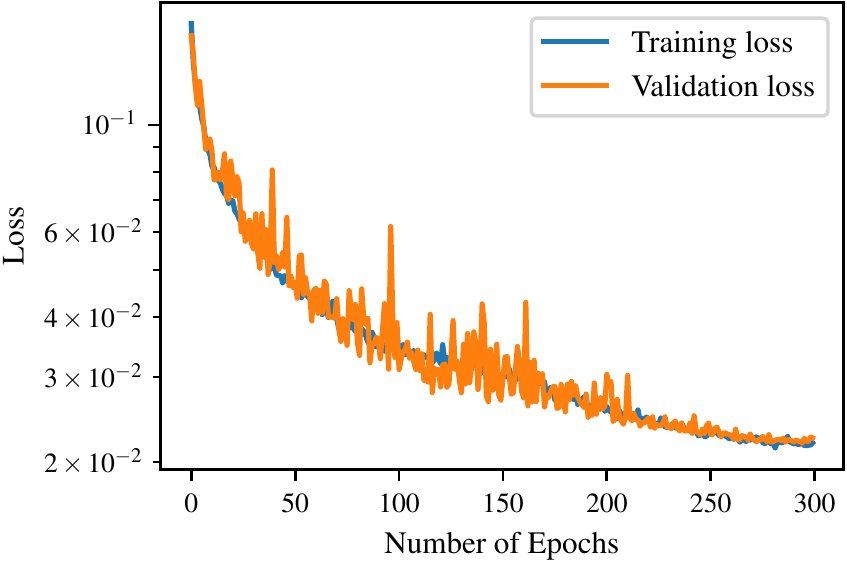}\\
    \includegraphics[width=\hsize]{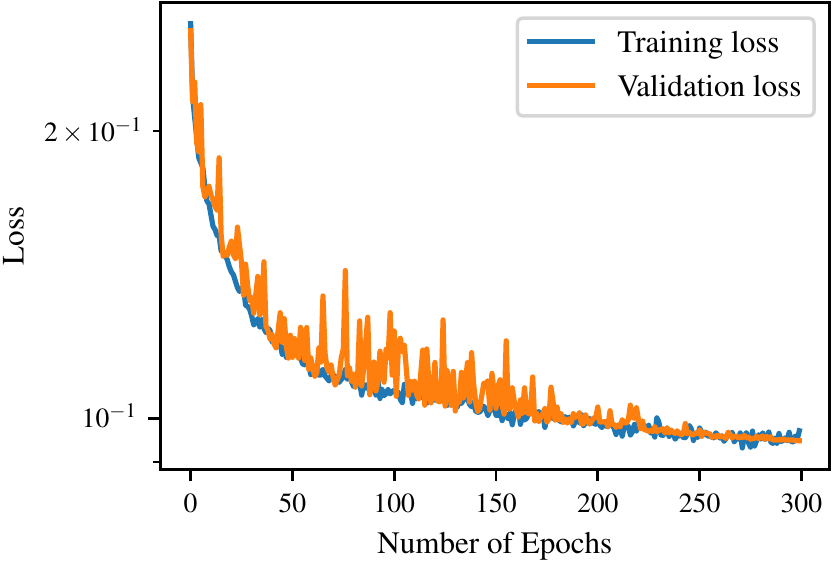}
    \caption{Loss curves for the training process of the network with noiseless (top) and noisy (bottom) input data. Loss values as a function of epoch are shown separately for training and validation data.}
    \label{fig:loss}
\end{figure}

The loss curves shown in \autoref{fig:loss} illustrate the learning process of the network without noise (top) and with added noise (bottom). 
After a sharp drop at the beginning of the training, a period with more spikes in the validation loss occurs. However, this has no negative impact on the reconstruction quality of the final network since the training process continues smoothly for the last 100 epochs.
Both curves show a similar behavior for the individual training sessions, meaning that the training converges well even for noisy input data. %The effects of correlated noise will be investigated in a forthcoming paper.

\section{Model Evaluation}
\label{sec:reconstruction}

\begin{figure*}
    \centering
    \resizebox{\hsize}{!}
    {\includegraphics[width=\hsize]{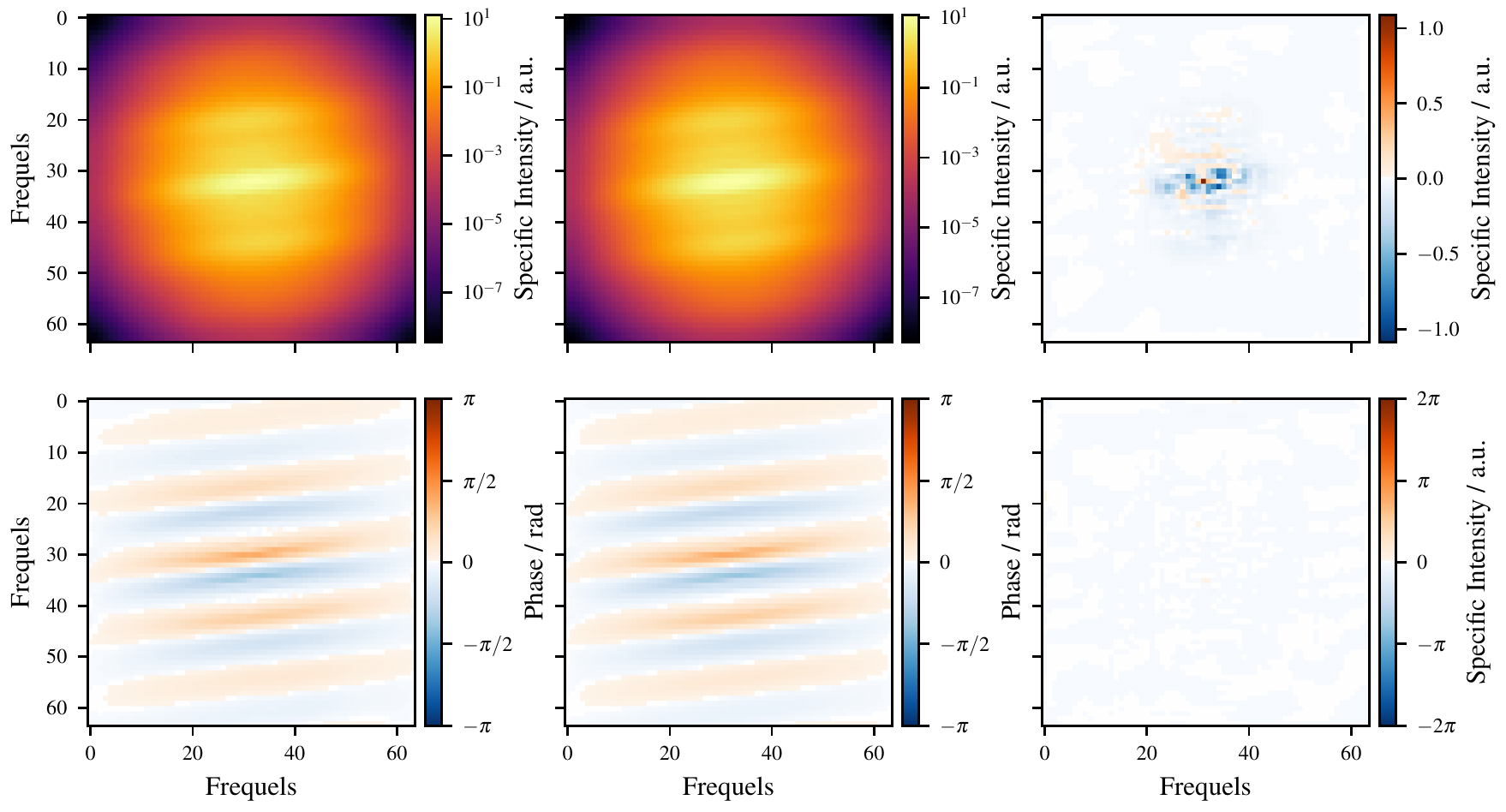}}
    \caption{Exemplary reconstruction for the training session with noiseless input data $V_\text{noiseless}$. Visualization of prediction (left), true distribution (middle) and the difference between both (right). Results are shown for amplitude (top) and phase (bottom).}
    \label{fig:amp-phase-clean}
\end{figure*}

\begin{figure*}
    \centering
    \resizebox{\hsize}{!}
    {\includegraphics[width=\hsize]{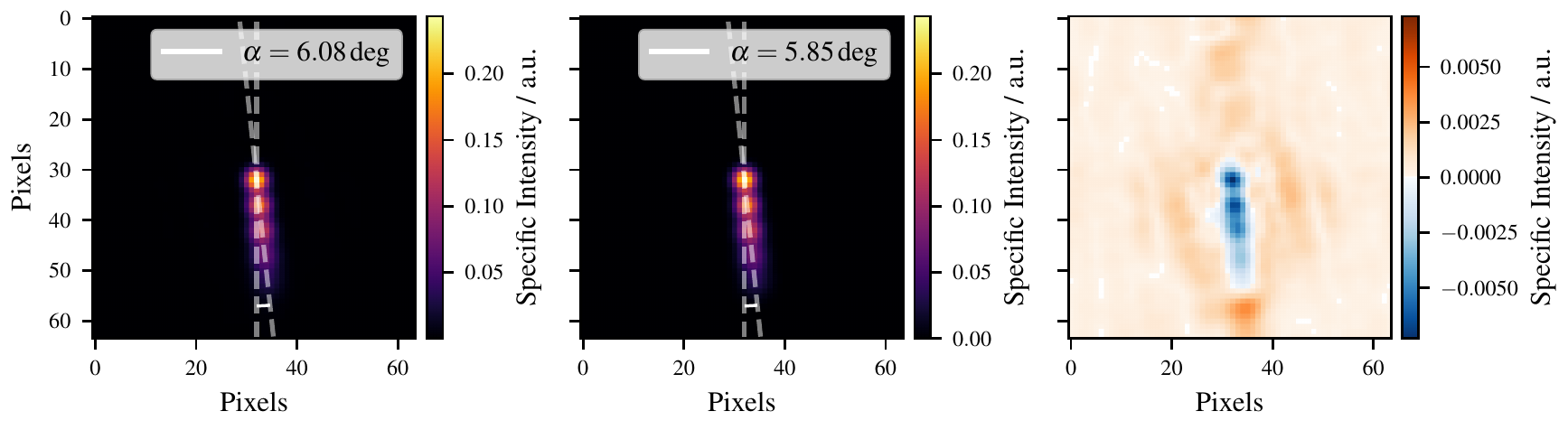}}
    \caption{Source reconstruction with the predicted amplitude and phase distributions for noiseless input data $V_\text{noiseless}$. Resulting clean image (left), simulated brightness distribution (middle) and difference between both (right). The jet angle $\alpha$, which was calculated using a PCA, is given for both source images.}
    \label{fig:source-recon-clean}
\end{figure*}

In this section, we test the ability of the trained models to reconstruct incomplete \uv data. We apply the models described in \autoref{sec:training} to images from a dedicated test data set. 
Next, we calculate the deviation of reconstructed amplitude and phase from the true distributions. Then we compare the reconstructed source images, which are generated by the inverse Fourier transformation, with the true images. %Applying more advanced evaluation methods serves to assess the results in more detail. These illustrate the precise reconstruction of the sources' brightness distributions.
Finally, we investigate the reconstructed jet angles, source areas, and the specific intensity of the core components are evaluated.

\begin{figure*}
    \centering
    \resizebox{\hsize}{!}
    {\includegraphics[width=\hsize]{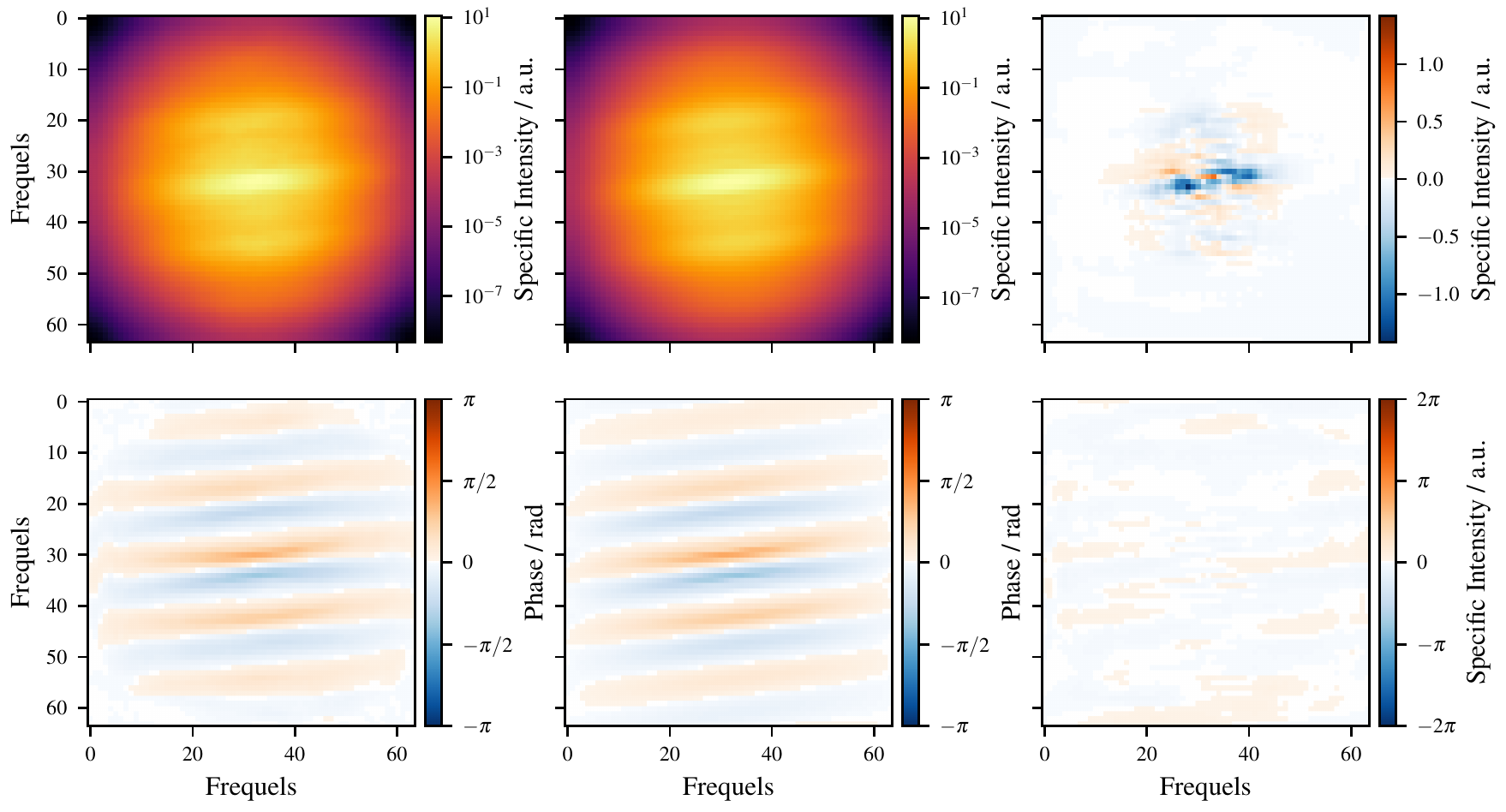}}
    \caption{Exemplary reconstruction for the training session with noisy input data $V_\text{noisy}$. Visualization of prediction (left), true distribution (middle) and the difference between both (right). Results are shown for amplitude (top) and phase (bottom).}
    \label{fig:amp-phase-noise}
\end{figure*}

\begin{figure*}
    \centering
    \resizebox{\hsize}{!}
    {\includegraphics[width=\hsize]{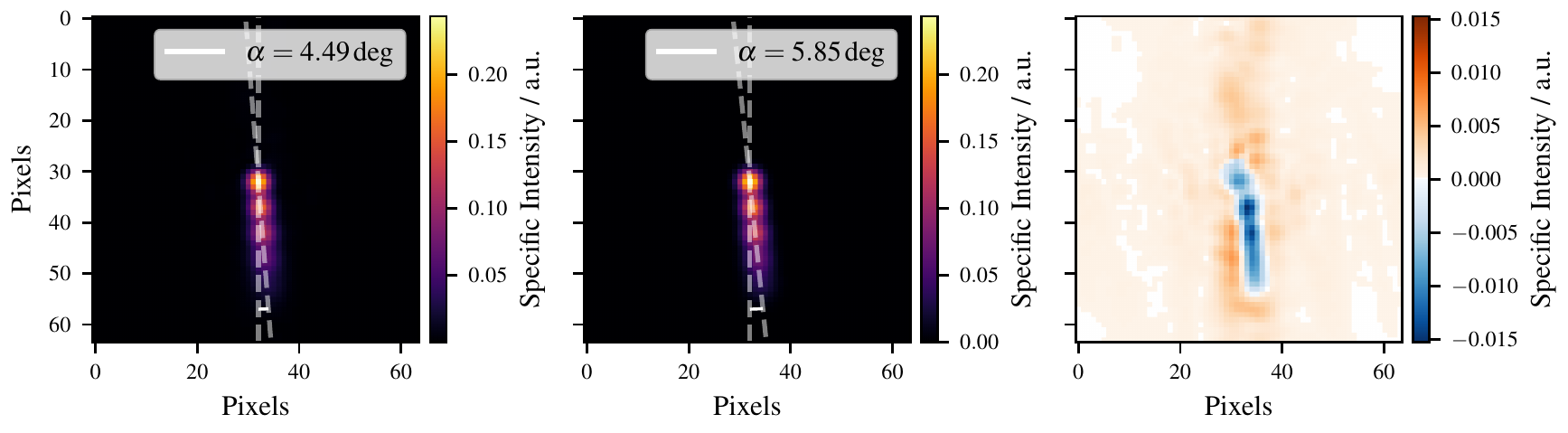}}
    \caption{Source reconstruction with the predicted amplitude and phase distributions for noisy input data $V_\text{noisy}$. Resulting clean image (left), simulated brightness distribution (middle) and difference between both (right). The jet angle $\alpha$, which was calculated using a PCA, is given for both source images.}
    \label{fig:source-recon-noise}
\end{figure*}

\subsection{Reconstruction of Fourier Data and Brightness Distributions}

The Deep Learning models allow for the reconstruction of missing information in incomplete Fourier data. 
Their execution time is of the order of milliseconds per image, which, compared to cleaning algorithms implemented in standard software such as DIFMAP or WSCLEAN, constitutes a significant speed advantage.
%When using established cleaning software, it can take days to a week to generate the final clean image, as the cleaning parameters have to be adjusted in iterative runs.
%In the case of analysis using Deep Learning techniques, the clean image can be generated in a day, even considering the long training process of the model.
%The Deep Learning model can be built in one long training session, which is then only fine-tuned for individual observations in much shorter training sessions.

For a first assessment of the reconstructions, we compare the reconstructed input data to the true distributions. By calculating the difference, we can identify areas with reconstruction problems. We summarize the results in \autoref{fig:amp-phase-clean}. In these figures, the first row shows the amplitude and the second row the phase. We use the predicted \uv spaces to create reconstructed source images, called clean images, by applying the inverse Fourier transformation. The comparison with the simulated brightness distribution provides a first indication of the quality of the reconstruction and helps to identify areas with reconstruction problems. Finally, we calculate the jet angle, $\alpha$, using a Principal Component Analysis (PCA). In PCA, a new basis is searched within the data. The goal is to maximize the information contained when projecting onto this basis. In the case of radio galaxy images, the axis through the jet forms a basis which can then be used to determine the jet angle. This is accomplished by computing the covariance matrix of the image and determining the eigenvalues and vectors of this matrix. The searched angle is then calculated by a tangent relation between the two eigenvectors. 

To investigate the model's ability to handle noise, we evaluate the results of two different training sessions. The first session was trained with noiseless input data, the second session was trained with noisy data. 
To compare our model to established imaging software, we perform the analysis of a dedicated data set using \texttt{wsclean}.
In the following, we present results produced with the example source shown in \autoref{sec:simulations} 

\subsubsection{Noiseless Input Data}\label{sec:results_noiseless}

\autoref{fig:amp-phase-clean} shows the reconstructed amplitude and phase for a training session with noiseless input data and different sampling masks. Overall, there is good agreement between prediction and truth for, both, amplitude and phase, except for small areas in the central part of the amplitude map. The trained model even enables the reconstruction of small-scale structures.
The deviations, i.e. the difference between prediction and truth, confirm this.
The agreement between prediction and 
truth can also be seen in the clean image in \autoref{fig:source-recon-clean}.
At first glance, one can hardly detect any differences between estimated and true brightness distribution. Calculating the difference between both reveals a small underestimation of the specific intensity in the source region. Additionally, artificial structures parallel to the central source region appear. The good reconstruction of amplitude and phase leads to differences in the two brightness distributions that are an order of magnitude smaller than the simulated specific intensity. The remaining background structures can be further constrained by comparing the contours of prediction and truth, visible in \autoref{fig:contour-plot-clean}. In this image, the red boundary contains all specific intensity above \SI{10}{\percent} of the source's peak intensity. Since this boundary does not include any of the background structures, the missing specific intensity is smaller than \SI{10}{\percent} of the source's peak intensity. The source area defined in this way is in good agreement with the true source area, with a ratio of \num{0.98}. 
The jet angles for prediction and truth are also consistent as their deviation is \SI{0.23}{\degree}. 
%In conclusion, the overall good reconstruction ability of the trained model is also apparent for the advanced evaluation methods presented in \autoref{subsec:advanced-eval-methos}.

\subsubsection{Noisy Input Data} \label{sec:noisy_input}

\autoref{fig:amp-phase-noise} shows the prediction for the same example described above, but with noisy input data. It is evident that the results do not differ much from the results with noiseless input data. The predicted amplitude misses some more nuances in the center of the image. The reconstruction of the small-scale structures still works works reliably. 
For the phase, deviations between prediction and truth become larger in the entire image. 
Therefore, noisy input data leads to losses in the reconstruction of the small-scale structures. Still, the differences between both distributions are relatively small. 
This result can be seen
in the reconstructed clean image shown in \autoref{fig:source-recon-noise}. Again, a visual comparison between reconstructed and true brightness distribution does not show large deviations. The calculation of the difference reveals a small underestimation of the specific intensity in the central part of the source. 
% Furthermore, artificial structures at the end of the jets become visible. 
The deviations are about an order of magnitude smaller than the simulated specific intensities.
The good agreement between prediction and truth is also evident in \autoref{fig:contour-plot-noise}. The outer red boundary shows that the background structures make up less than \SI{10}{\percent} of the source's peak intensity. Not only are the images well reconstructed, but noise added to the input data is cleaned up in the reconstruction process.
The calculated jet angles for prediction and truth are also consistent as their deviation is \SI{1.37}{\degree}. 
%In conclusion, the overall robust and good reconstruction results confirm the ability to process noisy input data of the developed model.

To further increase the complexity of our data and to simulate possible measurement effects in \uv{space}, we add additional white noise directly to the visibilities. 
For this, random values are drawn from a Gaussian distribution with a mean of 0 and a standard deviation of 0.05. Afterwards, the random sample is added to the real and the imaginary part of the data in Fourier space to create noise corrupted visibilities $V_\text{noisy\&white noise}$, as described in \autoref{sec:noise}.
% Thus, we can create a data set corrupted with noise added in image space, as described in \autoref{sec:noise}, and with additional noise corruption in visibility space. 
Due to the characteristics of the Fourier transformation, both types of noise result in distinct artifacts in the dirty images, see \autoref{fig:recons_source}. The white noise in \uv space leads to a sensitivity limit below which the information about the source area is lost. With the selected settings this sensitivity limit is a specific intensity of \num{5.47e-5}.

\subsubsection{Processing with \texttt{wsclean}}
\label{sec:wsclean}

\begin{table}[ht!]
    \centering
    \caption{Overview of the parameter settings utilized to create the clean images using \texttt{wsclean}.}
    \begin{tabular}{lr}
    \toprule
    Parameter & Setting \\
    \midrule
    size & \SI{64}{pixels} \\
    scale & \SI{0.39}{masec} \\
    mgain & \num{0.3} \\
    gain & \num{0.005} \\
    niter & \num{50000} \\
    \bottomrule
    \end{tabular}
    \label{tab:wsclean-settings}
\end{table}

\begin{figure*}
    \centering
    \includegraphics[width=\hsize]{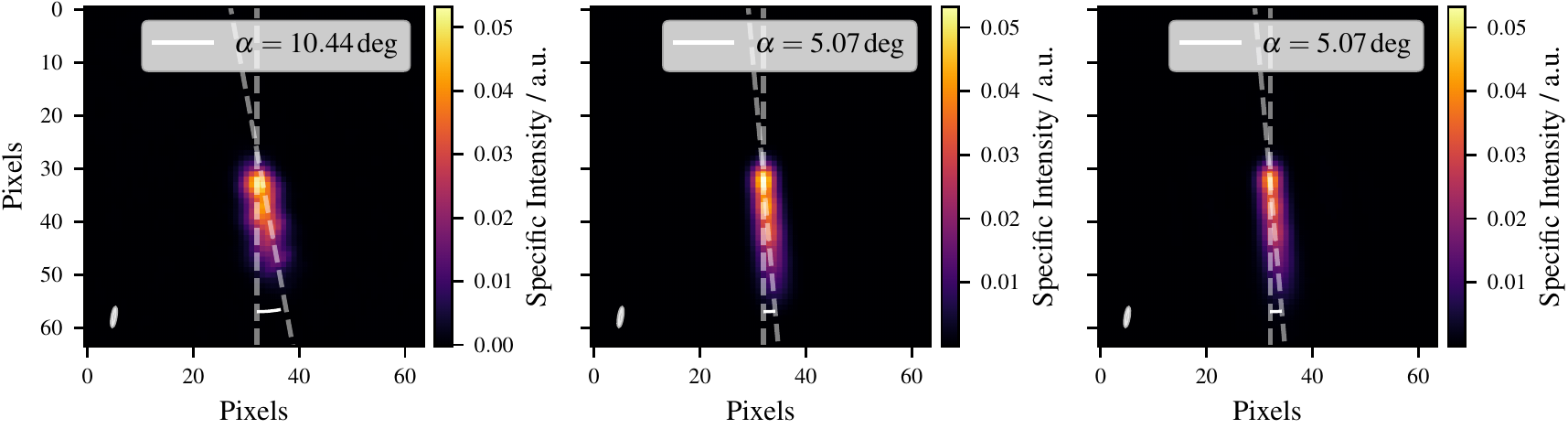}
    \caption{Clean image of a source with an one-sided jet created using \texttt{wsclean} (left) and simulated brightness distribution of the source convolved with the clean beam calculated via \texttt{wsclean} (middle). For comparison, a reconstruction of the same source using our Deep Learning approach trained with noiseless input data is shown (right), which is also convolved with the clean beam calculated via \texttt{wsclean}.} The full-width half-maximum clean beam sizes are shown on the lower left.
    \label{fig:one-wsclean}
\end{figure*}

\begin{figure*}
    \centering
    \includegraphics[width=\hsize]{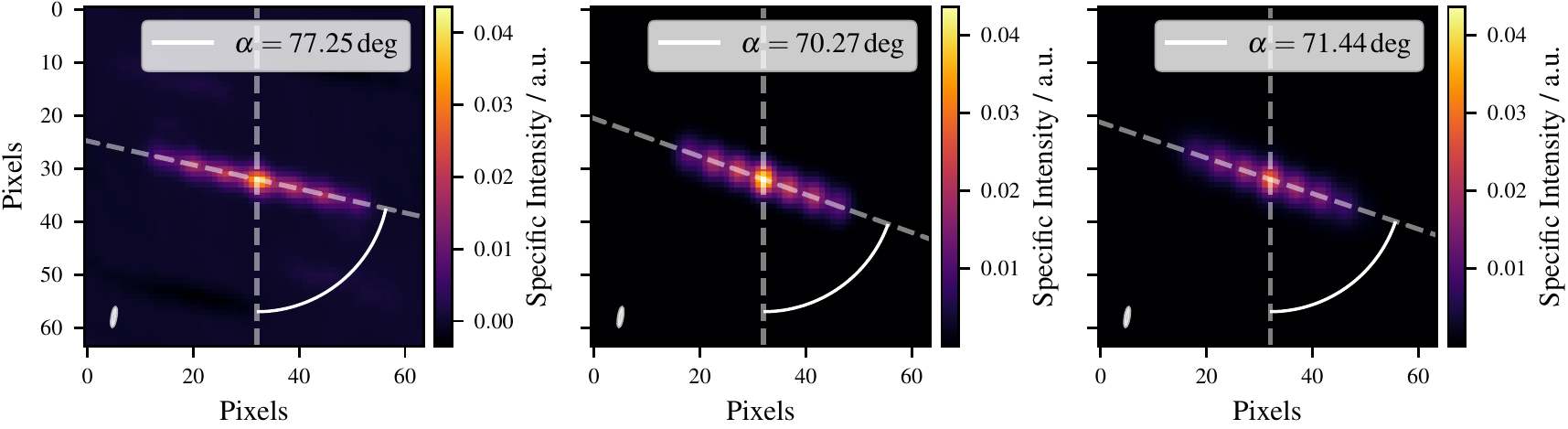}
    \caption{Clean image of a source with a two-sided jet created using \texttt{wsclean} (left) and simulated brightness distribution of the source convolved with the clean beam calculated via \texttt{wsclean} (middle). For comparison, a reconstruction of the same source using our Deep Learning approach trained with noiseless input data is shown (right), which is also convolved with the clean beam calculated via \texttt{wsclean}.} The full-width half-maximum clean beam sizes are shown on the lower left.
    \label{fig:two-wsclean}
\end{figure*}

In order to compare our reconstructions against data that have been cleaned using established imaging methods, we create a stack of clean images using \texttt{wsclean}.
To this end, we use the same simulated brightness distributions, as described in \autoref{sec:simulations}. 
As \texttt{wsclean} uses measurement sets as input format, the second part of the simulation had to be adjusted.
Firstly, we can no longer use a sampling mask to mimic an observation since ungridded visibilities are needed as input data.
To calculate these visibilities for each baseline, we use a simple \texttt{RIME} formulation like the one described in \cite{rime}.
Here, we only apply the direction-independent phase delay term for a better comparability with the input data for the neural network models.
The phase delay term corresponds to the Fourier kernel and does not introduce any further noise corruption.
Thus, the following equation for determining the uncorrupted visibility of the antenna pair $pq$ is obtained:
\begin{align}
      V_{pq} &= \sum_{l} \sum_{m} K_p(l, m) B(l, m) K_q^H(l, m),
\end{align}
with the source brightness $B(l, m)$ and the phase delay Jones matrix $K(l, m)=\exp(-2\pi i[ul + vm])$.
Here, $u$ and $v$ describe the coordinates of the current baseline corresponding to antenna pair $(p,q)$ in direction cosines. The $w$-term is neglected as $\sqrt{1-l^2-m^2} \approx 1$ is valid in our case.
Furthermore, $H$ denotes the conjugate transpose operation.
In order to ensure that the intensities of the visibilities are comparable with the flux density of the simulated brightness distribution, the summed flux density of the images is normalized to one.

In the next step, the simulated complex visibilities are written to \texttt{FITS} files. %Therefore, we developed a \texttt{FITS} writer following the descriptions in AIPS Memo 117 (revised).
Afterward, we use the \texttt{casa importuvfits} task to convert to measurement set format.
The obtained data is fed to \texttt{wsclean} to create clean images utilizing the cleaning parameters summarized in \autoref{tab:wsclean-settings}.

\autoref{fig:one-wsclean} shows a clean image of a source with a one-sided jet generated using \texttt{wsclean} (left) and the corresponding simulated brightness distribution of this source convolved with the clean beam calculated by \texttt{wsclean} (middle).
The comparison of both images shows that the general structure of the source is well reconstructed in the clean image.
However, in this example the flux density is slightly overestimated as the reconstructed source is more blurred. Especially in the core region this becomes obvious.
Besides the difference between the simulated and the reconstructed viewing angle, the reconstruction of the individual jet components works well in general.
For reference, the reconstruction of our Deep Learning approach is shown on the right, which for a better comparison is convolved with the clean beam calculated via \texttt{wsclean}.
Also in the case of two-sided jets \texttt{wsclean} is able to reconstruct individual components.
The example in \autoref{fig:two-wsclean} illustrates that the jet structure in the clean map (left) is reconstructed well. However, there is a difference between the simulated and the reconstructed jet angle. Additionally, faint background artifacts are visible, which occur more frequently in the case of two-sided jets. This is due to the fact that it is impossible to find a set of cleaning parameters for $10\,000$ sources that gives optimal results for all sources. The results show that, in the case of one-sided sources, the used parameter set leads to more accurate results.
Furthermore, the flux density of the reconstructed source is slightly underestimated.
This can be caused by several reasons:
First, the cleaning might not have been performed deeply enough.
In this case, tuning of cleaning parameters can be a possible solution.
%, but is not feasible for large data sets like the ones collected by sky survey.
Secondly, the difference in flux density may be caused by the large information loss during sampling.
Especially in the central region of the \uv plane, many data points are missing. These data points contain a lot of information about the large-scale brightness distribution of the source image.
% Still, the general source morphology can be reconstructed well, which is also confirmed by the small deviation of \SI{2.184}{\degree} between simulated and estimated jet angles.
% Again, the reconstruction using Deep Learning is shown (right), which is convolved with the clean beam calculated by \texttt{wsclean} for a better comparison.

\subsection{Advanced Evaluation Methods}
\label{subsec:advanced-eval-methos}

\begin{figure}
    \centering
    \includegraphics[width=\hsize]{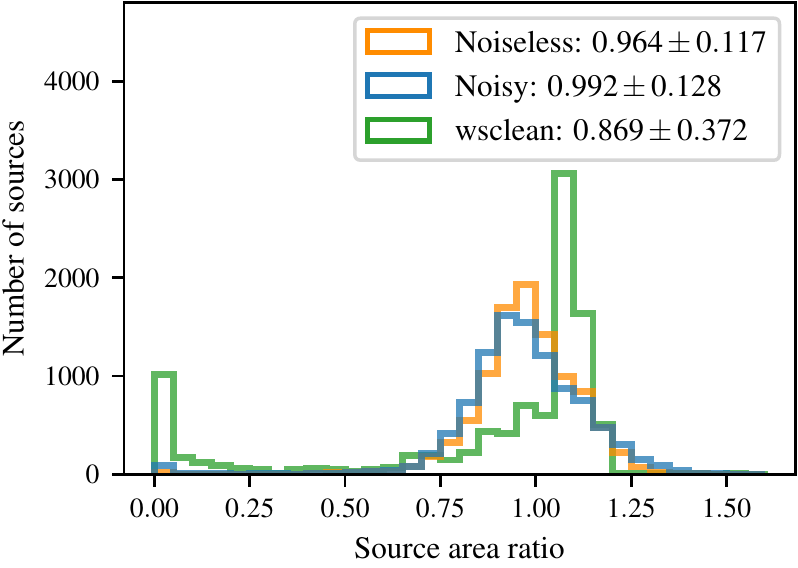}
    \caption{Histogram of the source area ratios between prediction and truth. 
    Results for a Deep Learning model trained with noiseless (orange), one trained with noisy input data (blue) and clean images generated with \texttt{wsclean} (green) are displayed.
    It becomes clear, that the distributions match well, which is supported by the mean and the standard deviation. In the case of \texttt{wsclean} a small overestimation of the area is visible.}
    \label{fig:hist-area-noise}
\end{figure}

\begin{figure}
    \centering
    \includegraphics[width=\hsize]{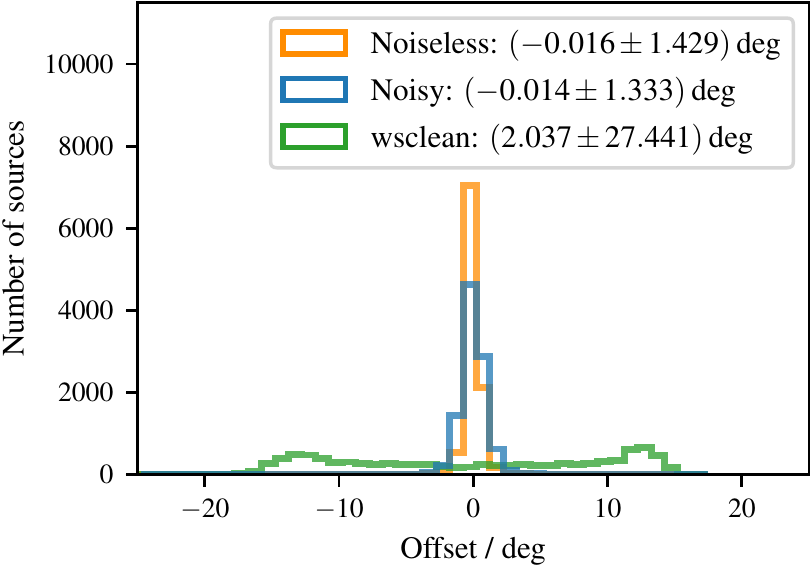}
    \caption{Histogram of the jet offsets 
    for a Deep Learning model trained with noiseless (orange), one trained with noisy input data (blue) and clean images generated with \texttt{wsclean} (green). The offset range is capped from -\SI{25}{\degree} to \SI{25}{\degree} for visibility reasons.
    Only small differences are present between the two distributions obtained from the Deep Learning models.
    Reconstructed jet orientations fit the true values well,
    which is supported by the means and standard deviations.
    Small pixel offsets in the cleaning process can already cause large offsets of the jet angle.
    This is a possible explanation for the larger deviations occurring in the case of \texttt{wsclean}. 
    }
    \label{fig:jet-offsets-combined}
\end{figure}

\begin{figure}
    \centering
    \includegraphics[width=\hsize]{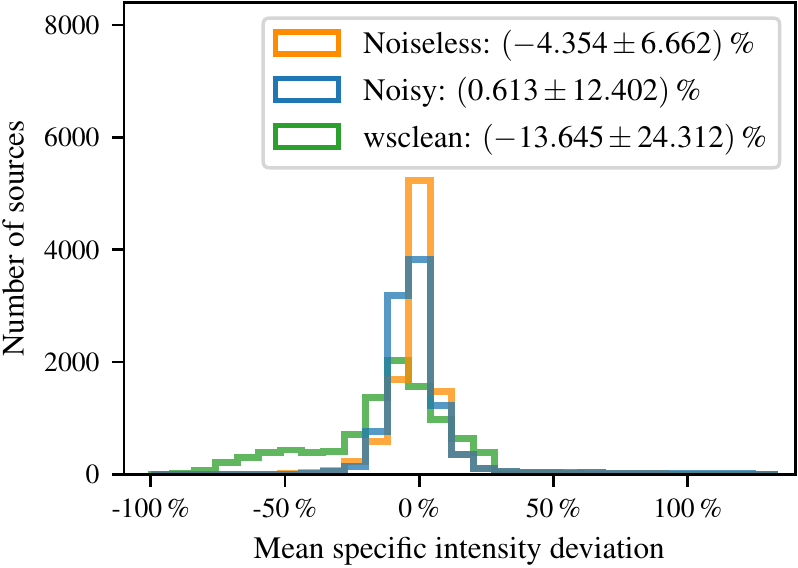}
    \caption{Histogram of the mean specific intensity deviation in the core component for a Deep Learning model trained with noiseless (orange), one trained with noisy input data (blue) and clean images generated with \texttt{wsclean} (green). The results are near the optimal value of zero on a similar level, although there are some outliers, which is represented by the relatively high standard deviation.
    For \texttt{wsclean} the peak is shifted slightly into the negative values, which indicates a slight underestimation of the flux densities.}
    \label{fig:hist-mean-diff}
\end{figure}

In this section, we compare the reconstructions of the Deep Learning models with results using \texttt{wsclean}. For Deep Learning, the sampled input data are reconstructed in Fourier space. Subsequently, the clean images are created using the inverse Fourier transform. In the case of \texttt{wsclean}, the visibilities stored in measurements sets are used to create the clean images. Thus, two dedicated test data sets are created each consisting of $10\,000$ images.

In order to compare the reconstructed and true source areas, we set an outer source boundary of \SI{10}{\percent} of the source's peak intensity. 
Afterwards, we measure the source areas using Leibniz' Sector formula \citep{leibniz}, which provides a relation between the path integral and the area of an enclosed region.
The enclosure of the source areas is obtained by calculating the contour levels of the source using \texttt{matplotlib}.
Then, the path integral is numerically approximated. The resulting values are used to calculate the source areas with the help of the Leibniz' Sector formula.
Then, we can compute the area ratio between the reconstructed and the true brightness distributions. In the ideal case, the area ratio is one. Smaller values correspond to an underestimate of the reconstructed source area. Larger predicted source areas lead to ratios above one. With this method, we obtain an estimate of the number of background artifacts. Images with reconstruction errors show more background artifacts and less specific intensity in the source region at the same time. Therefore, they have area ratios below one. \autoref{fig:hist-area-noise} shows the area ratios for the complete test data set for the Deep Learning model trained with noiseless input data (orange), the Deep Learning model trained with noisy input data (blue) and for the images created with \texttt{wsclean} (green). All three distributions peak around the optimal value of one. A small deviation between predicted and true source areas is confirmed by the mean and standard deviation of \num{0.964\pm0.117} (noiseless) and \num{0.992\pm0.128} (noisy).
The smearing to values above 1 on the right side of the peak in the case of \texttt{wsclean} indicates that the area of sources that can be reconstructed is overestimated.
Some source distributions cannot be reconstructed by any of the models, which is represented by the peaks at a ratio of zero.
These instances occur more frequently in the case of \texttt{wsclean}.
This results in mean and standard deviations of \num{0.869\pm0.372} (\texttt{wsclean}).
Both reasons lead to a standard deviation of the distribution obtained with \texttt{wsclean} that is 3 times larger than that of the Deep Learning results.
For all distributions, the optimal value is within the uncertainty boundaries. Comparing the two Deep Learning models, the mean values for clean and noisy input data differ by only 0.028, which indicates that the architecture can reach the same reconstruction level for both input types.

In a second experiment, we investigated the reconstruction of the jet angles. In \autoref{fig:jet-offsets-combined}, three histograms containing the differences between the predicted and the true jet angles of $10\,000$ test images are shown. The orange histogram represents the Deep Learning model trained with noiseless input data, while the blue histogram represents the Deep Learning model trained with noisy input data. The results are similar, with the reconstruction using noisy input data faring slightly better, which is supported by the mean and standard deviation of \SI{-0.016\pm1.429}{\degree} (noiseless) and \SI{-0.014\pm1.333}{\degree} (noisy) for the distributions.
In both cases, there are only slight deviations between the jet angles calculated from the reconstructed and the true images, which is shown by the small standard deviations. 
As one can see, the reconstruction with noisy input data performs slightly better than the one with noiseless input data.
% but the latter also performs well on those input data that are more difficult to reconstruct.
The distribution of the jet angle offsets obtained from the images cleaned using \texttt{wsclean} is shown in green.
It is evident that there are only a few cases where a deviation of more than 15 degrees occurs.
Nevertheless, two maxima appear at offsets of around \SI{-12}{\degree} and \SI{12}{\degree}, which exceed the peak at zero in height.
This may be caused by a systematic problem in the reconstructions with \texttt{wsclean}.
Since in our analyses small image sizes are used, small deviations in the cleaned maps lead to large differences in the jet angle.
A deviation of one pixel can already lead to an offset of several degrees.
Additionally, we note that \texttt{wsclean} is not optimized for clean images of this size.
%hese facts have to be kept in mind when interpreting the jet offset distribution.
The evaluation of the complete test set results in a mean jet offset of \SI{2.037\pm27.441}{\degree} (\texttt{wsclean}).
Where the large standard deviation is caused by the two maxima at \SI{-12}{\degree} and \SI{12}{\degree}.

Finally, we compare the predicted and the true specific intensities of the core components. For this purpose, the Gaussian components of the simulated images are identified by the \texttt{blob detection} algorithm inside the \texttt{scikit-image} \citep{skimage} package. Afterward, the pixels with the brightest specific intensities, in our case the core component, are averaged for both truth and prediction, and then compared. In this way, the relative deviation of the mean true specific intensity in the first component can be computed. \autoref{fig:hist-mean-diff} visualizes the resulting values. The mean specific intensity deviation for the training session with noiseless input data (orange) and the training session with noisy input data (blue) is close to the optimal value of zero for both Deep Learning models. The standard deviation, however, is higher due to some outliers whose flux is over- or underestimated. %Nonetheless, this underlines the robustness of the architecture against uncorrelated noise.
In the case of \texttt{wsclean} the mean reconstructed flux densities are underestimated to a larger extent.
This is also reflected in the mean and standard deviation of the distribution which is \SI{-13.645\pm24.312}{\percent} (\texttt{wsclean}).
Again, reconstruction can be improved by tuning the cleaning parameters in \texttt{wsclean} for the different sources and we wish to reiterate that
\texttt{wsclean} is not designed to be used with the same settings on \num{10\,000} different samples.
%This fact results in the relatively poor performance on the whole test data set.

\begin{table*}[ht!]
    \sisetup{table-number-alignment = right}
    \centering
    \caption{Overview of the mean results of the three evaluation methods with clean, noisy and noisy plus additional white noise input maps and for input maps with different mask filling. A new training set is used for different noise models. The data sets consist of 50\,000 training images, 10\,000 validation images, and 10\,000 test images. For different sampling densities, we do not train a new model but create dedicated test data sets consisting of 10\,000 test amplitude and phase maps. The evaluation is done using the Deep Learning model trained on noisy input data.}
  \begin{tabular}{l
  S[table-format=-1.2]@{${}\pm{}$}S[table-format=1.2]
  S[table-format=-3.2]@{${}\pm{}$}S[table-format=2.2, table-number-alignment = left]
  S[table-format=2.2]@{${}\pm{}$}S[table-format=1.2, table-number-alignment = left]
  }
    \toprule
    & \multicolumn{2}{c}{Jet offset  [$\si{\degree}$]} & \multicolumn{2}{c}{Intensity deviation  [$\si{\percent}$]} & \multicolumn{2}{c}{Source area ratio} \\
    \midrule
    {Noiseless input $V_\text{noiseless}$} & -0.02 & 1.43 & -4.35 &  6.66 & 0.96 & 0.12 \\
    {Noisy input $V_\text{noisy}$}              & -0.01 & 1.33 &  0.61 & 12.40 & 0.99 & 0.13 \\
    {Noise \& white noise $V_\text{noisy\&white noise}$}    & 0.05 & 1.96 &  3.49 & 12.26 & 1.00 & 0.13 \\
    \midrule
    Sampling Density \\
    \midrule
    \SI{20}{\percent} & -0.07 & 3.14 & -8.99 & 16.26 & 0.85 & 0.19 \\
    \SI{50}{\percent} & 0.02 & 1.96 & 6.69 & 14.45 & 0.92 & 0.17 \\
    \SI{70}{\percent} & 0.01 & 3.12 & 11.94 & 63.60 & 0.91 & 0.17 \\
    \bottomrule
  \end{tabular}
  \label{tab:results_noise_filling}
\end{table*}

\begin{table*}[ht!]
    \centering
    \caption{Comparison of run-times to reconstruct the data with our trained \texttt{radionets} neural network model and \texttt{wsclean} for different image sizes. The programs are run for 100 times on one image. The presented values are the mean run-times with standard deviations.}
    \begin{tabular}{
    S[table-format=4]
    S[table-format=1.2(2)]
    S[table-format=1.4(4)]
    S[table-format=1.2(2)]
    }
    \toprule
    {Image size [pixels]} & {\texttt{radionets} run-time [$\si{\second}$]} & {Evaluation run-time [$\si{\second}$]} & {\texttt{wsclean} run-time [$\si{\second}$]} \\
    \midrule
    64 & 2.29\pm0.04 & 0.0031\pm0.0001 & 0.43\pm0.02 \\
    128 & 2.31\pm0.08 & 0.0081\pm0.0002 & 0.63\pm0.04 \\
    256 & 2.48\pm0.06 & 0.0269\pm0.0008 & 1.03\pm0.04 \\
    512 & 2.65\pm0.13 & 0.0995\pm0.0043 & 2.42\pm0.09 \\
    1024 & 3.20\pm0.06 & 0.4070\pm0.0052 & 9.00\pm0.34 \\
    \bottomrule
    \end{tabular}
    \label{tab:run-times}
\end{table*}

In order to evaluate the performance of the model applied to data affected by white noise, all three methods were applied to the new data set.
Mean and standard deviations of the resulting distributions are summarized in \autoref{tab:results_noise_filling}.
The reconstruction of the source area and the reconstruction of the jet angle do not change remarkably.
The reason that the source area  does not change that much despite the above-mentioned sensitivity limit is that only around 30 of our \num{10\,000} image test data set are below this limit.
In the case of the mean intensity deviation of the core component an increase of the mean and the standard deviation is visible.
This happens due to the sensitivity limit resulting from the white noise in \uv space.
Sources with particularly weak brightness distributions can no longer be reconstructed and appear as outliers in the distributions, leading to increased standard deviations.

\begin{figure}
    \centering
    \includegraphics[width=\hsize]{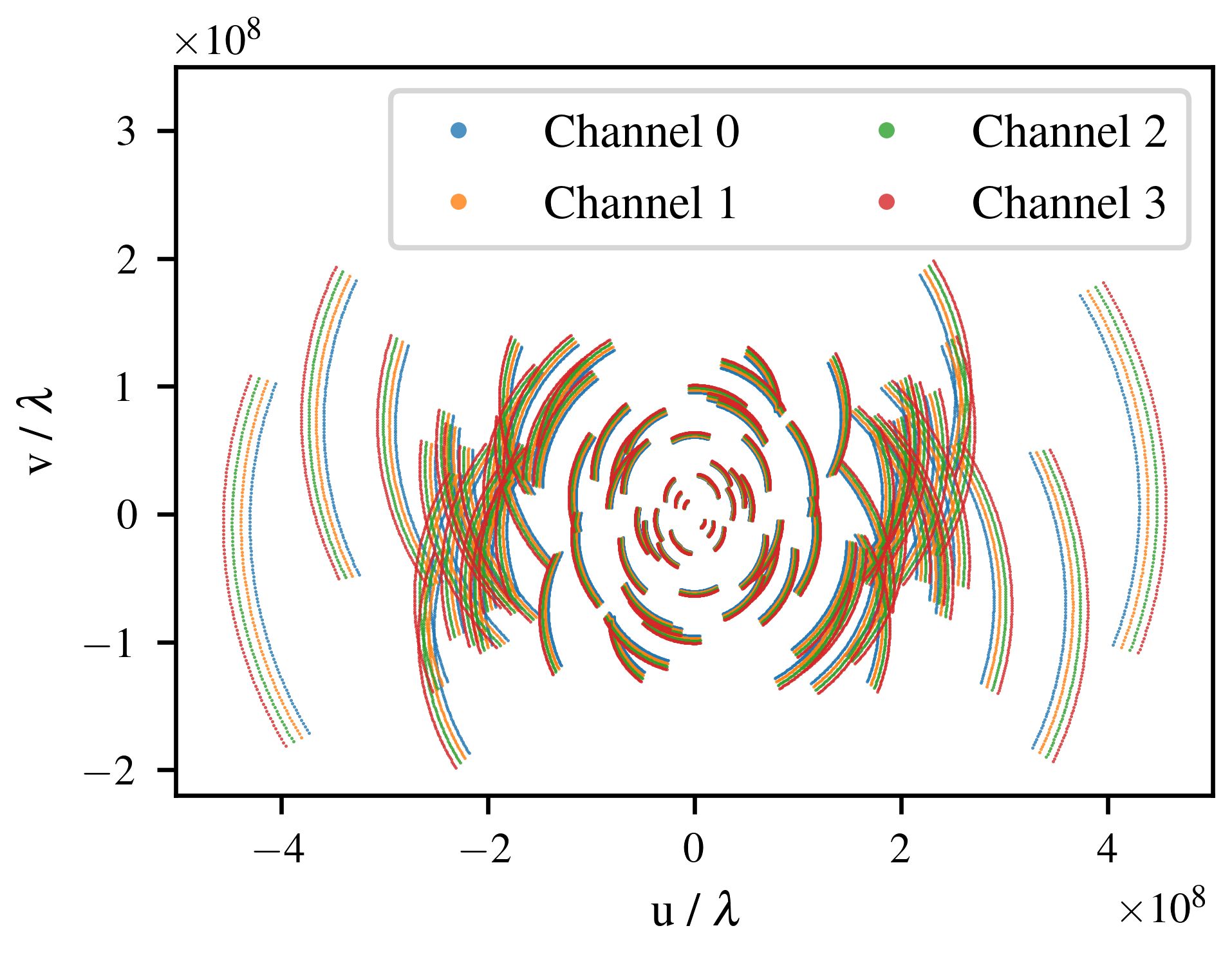}
    \caption{Exemplary \uv coverage for a simulated observation with 50 time steps and 4 frequency bands. The additional channels lead to a slightly improved coverage of amplitude and phase frequels. However, the greater advantage is the improved sensitivity of the simulation since the signal-to-noise ratio is improved by the increased number of data points.}
    \label{fig:multi_cahnnel}
\end{figure}

To test the dependence on the masks (i.e. \uv coverage), we evaluated three data sets with different fillings for the masks on the model trained with noisy input data (see \autoref{sec:reconstruction}).
The fillings are \SI{20}{\percent}, \SI{50}{\percent} and \SI{70}{\percent}. To achieve such high sampling densities we added the option to simulate multi-channel data. An example of a simulation with four channels is shown in \autoref{fig:multi_cahnnel}.
Here, the number of samples is increased by a factor of $4$. This results in more information collected per frequel, which increases the sensitivity of the simulated observation as the signal to noise ratio is improved.
\autoref{tab:results_noise_filling} summarizes the reconstruction results which were evaluated with the methods introduced in this chapter. Mean and standard deviations of the distributions show that the success of the reconstruction is correlated with the filling of the mask. 
% This is best illustrated by the mean, which decreases with increasing filling, just as expected. 
% This trend remains when non-interpolated data are reconstructed with our model, indicating that the quality of the nearest-neighbor interpolation is also linked to the sampling density.
%To be able to make an exact statement, we plan a detailed evaluation of the impact of the interpolation before the data is passed to the network in a future work.
The evaluation of the jet offsets and the mean intensity deviation illustrates that the reconstruction quality is greatly affected by small sampling densities. 
The values for the source area ratios and their standard deviations improve with larger sampling rates, as expected.
The behavior of the mean intensity deviations is opposite, as larger sampling densities lead to an overestimation of the specific intensity. This leads to larger values for mean and standard deviations. In future work, we will use a wider range of sampling rates in the training to further improve the robustness of our model.

To conclude, the sampling density is directly correlated with the model's ability to reconstruct the characteristics of the simulated jet. This confirms the robustness of our model, which made use of data with a sampling density of around \SI{30}{\percent}.

\subsection{Execution Times}
\label{sec:application-times}

Assuming that we can apply our network to a large number of similarly obtained data sets, the time for training the model becomes less important than the execution time on a single data set.
To evaluate this, we have summarized the run-times of our model and of \texttt{wsclean} for different image sizes in \autoref{tab:run-times}.
For the Deep Learning model, the run-time is given once for the whole \texttt{radionets} framework, which includes the loading of the model, the loading of the test data, and the saving of the clean image. 
Additionally, the pure execution time of the model is shown.
In order to determine the run-times of \texttt{wsclean}, the cleaning settings from \autoref{tab:wsclean-settings} were used.
The size parameter was adjusted here for the different image sizes.
% Additionally, it is to be noted that the \texttt{radionets} input data require odd pixel numbers.
% For \texttt{wsclean} the pixel sizes were increased by one as the implementation of the fast Fourier transformation works best for image sizes that consist of powers of two.
The results show that the \texttt{radionets} framework can generate clean images faster than \texttt{wsclean} for larger image sizes.
Considering the pure reconstruction time of the Deep Learning model, our models can reconstruct the input data faster for all image sizes.
%For this reason, our approach has a strong advantage when imaging large data sets.
\begin{figure*}
    \centering
    \resizebox{\hsize}{!}
    {\includegraphics[width=\hsize]{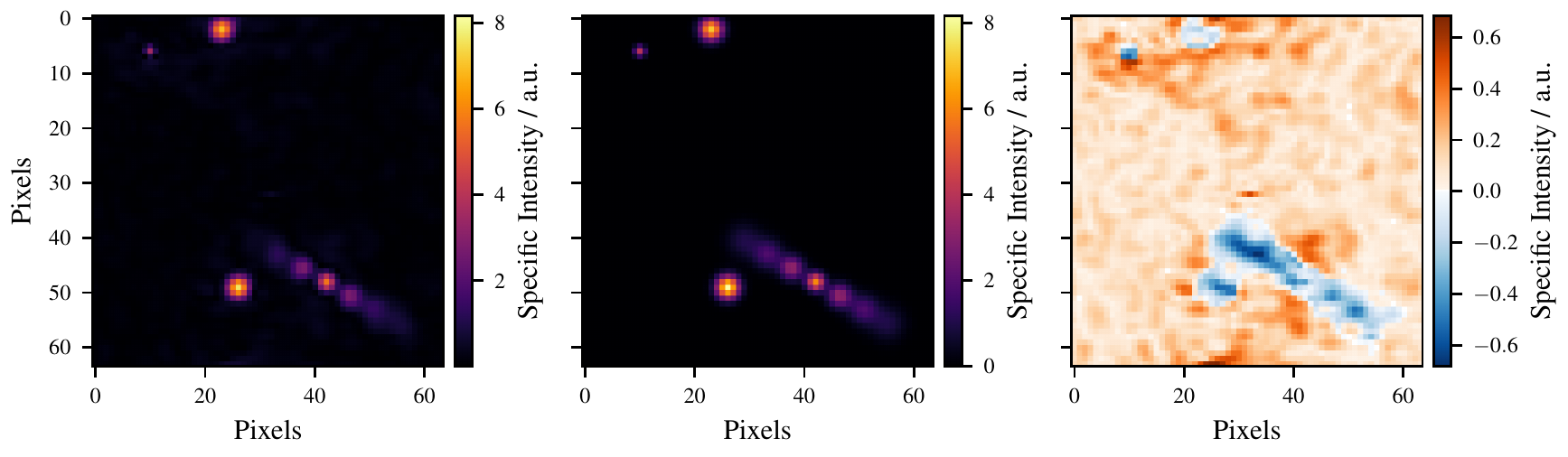}}
    \caption{Source reconstruction with the predicted amplitude and phase distributions for mixed data. Resulting clean image (left), simulated brightness distribution (middle) and difference between both (right).}
    \label{fig:mix-samp-pred}
\end{figure*}

\section{Further Analysis}
\label{sec:further_approaches}

\subsection{Improved Simulations}
\label{subsec:improved-simulations}

The current simulations mimic observations of radio galaxies built from Gaussian components.
%Incomplete \uv coverage generated using sampling masks are viable to demonstrate the feasibility of reconstructing incomplete Fourier data with neural networks.
More realistic simulations of radio interferometric observations require further adjustments.
The most important aspects are the beams of the individual telescopes inside the interferometer array, the effect of sidelobes, and the influence of Gaussian noise per baseline. In the future, we will directly simulate visibilities using the radio interferometer measurement equation (RIME) to describe the individual components of the observation \citep{rime} including the array responses.
Furthermore, it is modularly expandable to add effects of the ionosphere or noise corruption by telescope receivers.
Together with advanced radio sky simulations, this will provide the path to train Deep Learning networks that can be applied to real data.

First, we created a data set containing extended as well as point sources and trained a new network for 300 epochs with 50\,000 training and 10\,000 validation images. This has the advantage that we can test how the chosen architecture behaves on data that differs from what we have tested so far. To evaluate the performance of the network, we create a new test data set 
containing simulated point-like Gaussian sources with different sizes besides the extended radio galaxies.
Each image covers a randomly drawn number of these additional point-like sources between one and six.
The images have a size of $64 \times 64$ pixels and thus are comparable to sub-images of a larger sky survey.
Again, the methods described in \autoref{sec:simulations} are used to simulate observations with a radio interferometer.
In this way, we create a data set consisting of 10\,000 test images.
The sampled amplitude and phase distributions serve as input for the neural network.

In \autoref{fig:mix-samp-pred}, we show an example of a source reconstruction image that comes from the mixed data set. It is apparent that the positions of all point sources, even the ones with a very small specific intensities, are correctly reconstructed in the predicted image. The extended source in the lower right half of the image can also clearly be seen. For all sources, some intensity is missing as evident from the difference plot on the right side, but the maximal difference is just around $10\,\%$ of the maximal intensity. In addition, the background is predominantly reconstructed to zero, no major artifact can be seen. In summary, our network is able to reconstruct a mixed data set as well as a data set containing only extended sources.

\begin{figure}
    \centering
    \includegraphics[width=\hsize]{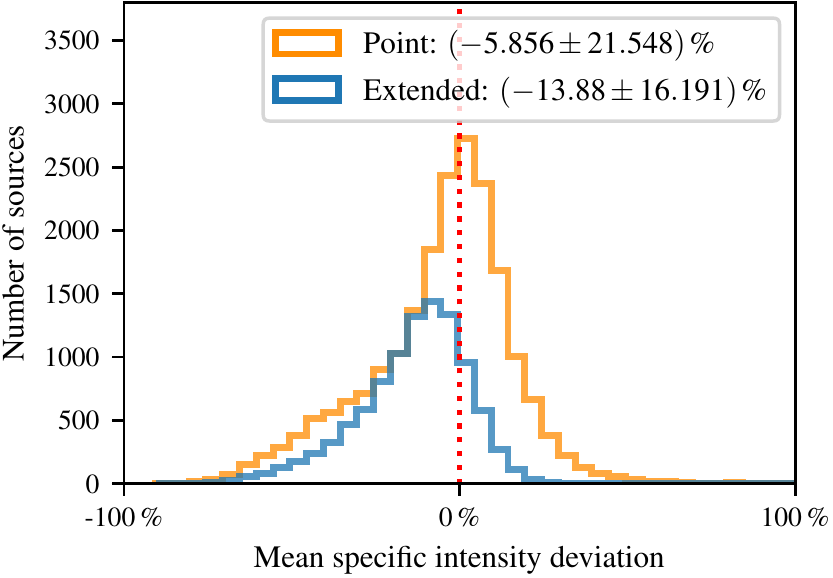}
    \caption{Histogram of the mean specific intensity deviation for the point-like and extended Gaussian sources contained in \num{10000} test images. The majority of the sources' flux is underestimated, as confirmed by the mean value of \SI{-5.865(21548)}{\percent} for the point sources and \SI{-13.88(16191)}{\percent} for the extended sources.}
    \label{fig:pointsources}
\end{figure}

For the evaluation of these mixed images, we compared the mean specific intensity of each source in the image in the same way as in \autoref{fig:hist-mean-diff}. For the extended sources, we summed up the specific intensity over the whole source area.
%, so that one flux value is compared for each predicted extended source. 
The results are summarized in a histogram shown in \autoref{fig:pointsources}, separately for point and extended sources. It turns out that most of the sources' intensity is underestimated, which is confirmed by the mean values of $\approx \SI{-6}{\percent}$ and $\approx \SI{-14}{\percent}$, respectively. We note that there are some outliers with large positive deviations but their numbers are very small.

\begin{figure}
    \centering
    \includegraphics[width=\hsize]{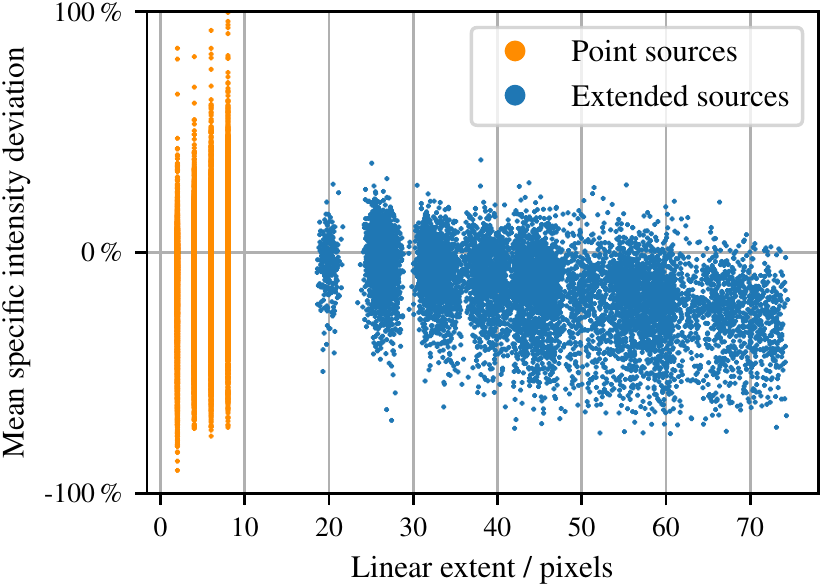}
    \caption{Relation between the linear extent and the mean specific intensity deviation. The orange values are point sources, the blue ones extended sources. The extended sources are underestimated, while there is not clear trend for the point sources. If \autoref{fig:pointsources} is included, it becomes clear that the majority of the sources is underestimated and only some outliers overestimated.}
    \label{fig:extent-deviation}
\end{figure}

Moreover, we evaluated a possible relation between the linear extent of the sources and the mean specific intensity deviation. The linear extent for the point sources is calculated via the standard deviation of the Gaussian kernel that was used to smear out the sources. For the extended sources, this was accomplished via the distance between the most distant blobs for an extended source plus the sigma values for these blobs to account for the spreading. The corresponding plot is shown in \autoref{fig:extent-deviation}. While the extended sources are generally underestimated, the point sources have more outliers in the overestimate region, but the majority is still underestimated, as \autoref{fig:pointsources} shows. Furthermore, we find no correlation between the linear extent and the intensity deviation.

\begin{figure}
    \centering
    \includegraphics[width=\hsize]{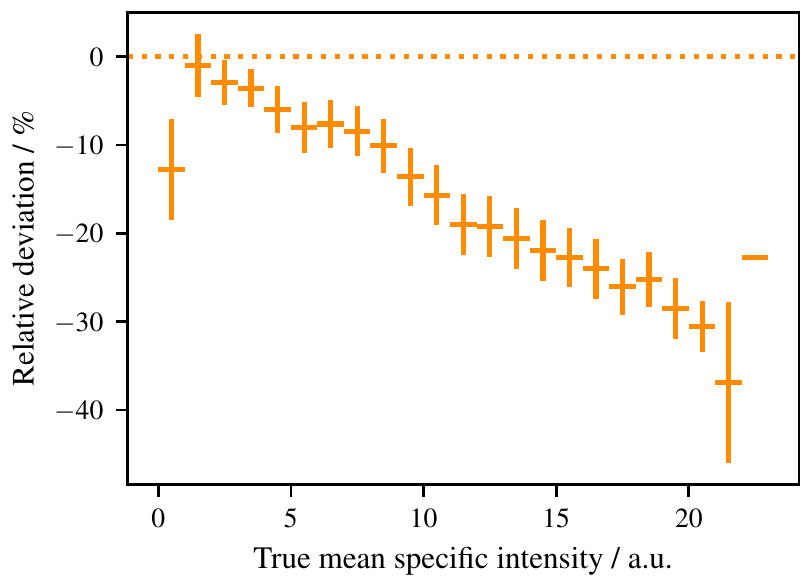}
    \caption{Relative deviation between predicted and true mean specific intensity shown for different intensity levels. The mean value is shown for the different intensity bins. Bin width is illustrated by the $x$ error bars. The uncertainty of the relative deviations is represented by the $y$ error bars.}
    \label{fig:2d-hist-mixed}
\end{figure}

For a more detailed look into the specific intensity, we show the relative deviations for the different intensities in \autoref{fig:2d-hist-mixed}. For small intensity values, the estimates are in accordance with the true values, except for the very smallest values. With increasing mean specific intensity the relative deviations increase, as expected from \autoref{fig:pointsources}. The uncertainty of the relative deviations are similar for all intensities, except for the smallest and the largest intensities.

\begin{figure*}
    \centering
    \resizebox{\hsize}{!}
    {\includegraphics[width=\hsize]{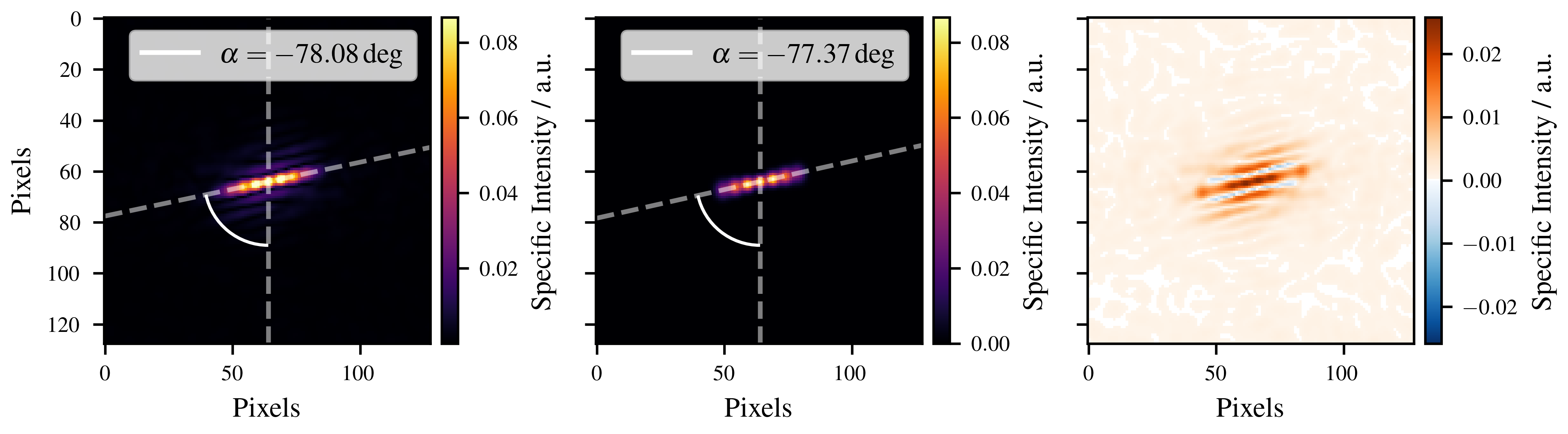}}
    \caption{Clean image resulting from the reconstructed amplitude and phase distributions generated by a basic network trained on ($64 \times 64$)-pixel maps without additional fine-tuning. Resulting clean image (left), simulated brightness distribution (middle) and difference between both (right). The jet angle $\alpha$, which was calculated using a PCA, is given for both source images.}
    \label{fig:dif_sizes_wo_fine_tune}
\end{figure*}

\begin{figure*}
    \centering
    \resizebox{\hsize}{!}
    {\includegraphics[width=\hsize]{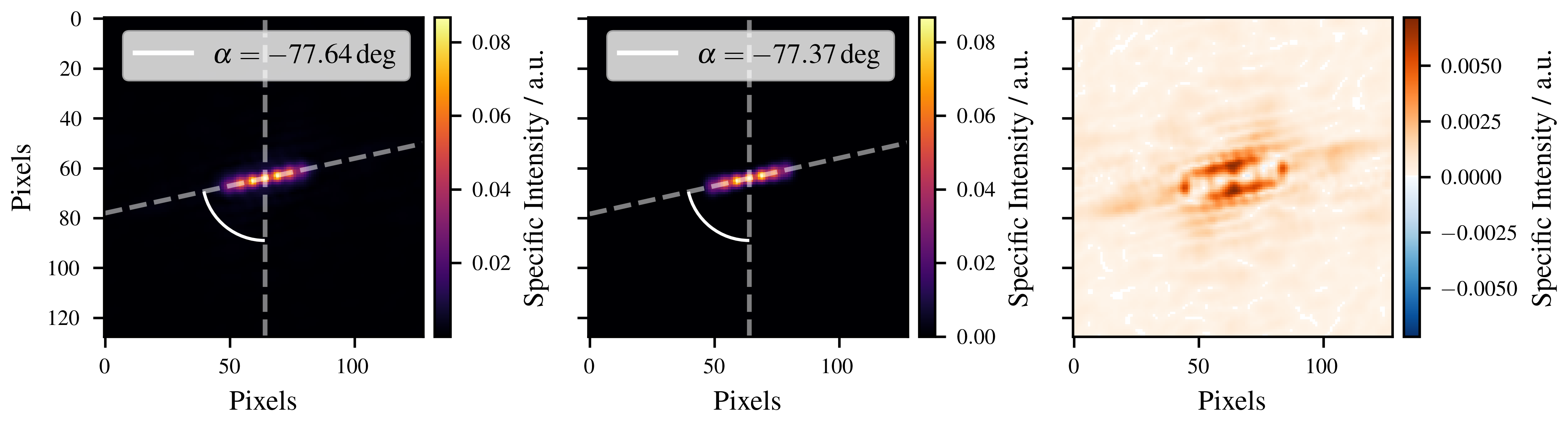}}
    \caption{Clean image resulting from the reconstructed amplitude and phase distributions generated by a basic network trained on ($64 \times 64$)-pixel maps, which was fine-tuned on the new data set with ($128 \times 128$)-pixel maps for another 20 epochs. Resulting clean image (left), simulated brightness distribution (middle) and difference between both (right). The jet angle $\alpha$, which was calculated using a PCA, is given for both source images.}
    \label{fig:dif_sizes}
\end{figure*}

In summary, our network is able to reconstruct a mixed data set containing point and extended sources on a similar level as the simple data set consisting of extended sources only. 
%This underlines the variability of our approach and encourages us to further expand and improve this approach in future projects.

\subsection{Independence of Image Size: Reconstruction of Larger Images with the same Architecture}
\label{subsec:image-size}

A feature of the proposed architecture is not to resize the input data, which means that the image size does not vary throughout the network, and thus any image size can be handled without the need for resizing or similar operations.

For illustration, we perform the training of a basic network on a data set with small-scale amplitude and phase maps, such as the maps presented in the previous sections.
Here, training on a big data set on short timescales is possible. Afterward, we fine-tune this network by training it with a data set consisting of larger amplitude and phase maps for another 20 epochs. This data set consists of $5\,000$ training maps and $1\,000$ validation maps. For these maps, we simulate a larger field of view and larger baselines at the same time, which results in higher spatial resolutions. The fine tuning takes around $\SI{30}{\minute}$.
This technique of transferring the learning process enables a fast convergence for the network to perform well on the new data set. The fine-tuning is necessary to adapt the new scaling in Fourier space. Without this, the individual source components show an increased specific intensity deviation and slight artifacts in the background occur, which can be seen in \autoref{fig:dif_sizes_wo_fine_tune}. Here, ($128 \times 128$)-pixel amplitude and phase maps were reconstructed using a basic network trained on ($64 \times 64$)-pixel maps. \autoref{fig:dif_sizes_wo_fine_tune} shows the clean image resulting by applying the inverse Fourier transformation to the reconstructed distributions. As comparison, \autoref{fig:dif_sizes} illustrates the clean image resulting from reconstructed ($128 \times 128$)-pixel amplitude and phase maps generated  with a network that received an additional fine-tuning for another 20 epochs on the new data set. After fine-tuning the intensity deviations and background artifacts decrease.
Furthermore, it is evident that the reconstruction quality barely differs from the reconstructions shown in \autoref{sec:reconstruction}.
This fact enables the possibility to train Deep Learning networks that can reconstruct data with larger image sizes without great effort when starting with a trained basic network.

\begin{figure*}
    \centering
    \resizebox{\hsize}{!}
    {\includegraphics[width=\hsize]{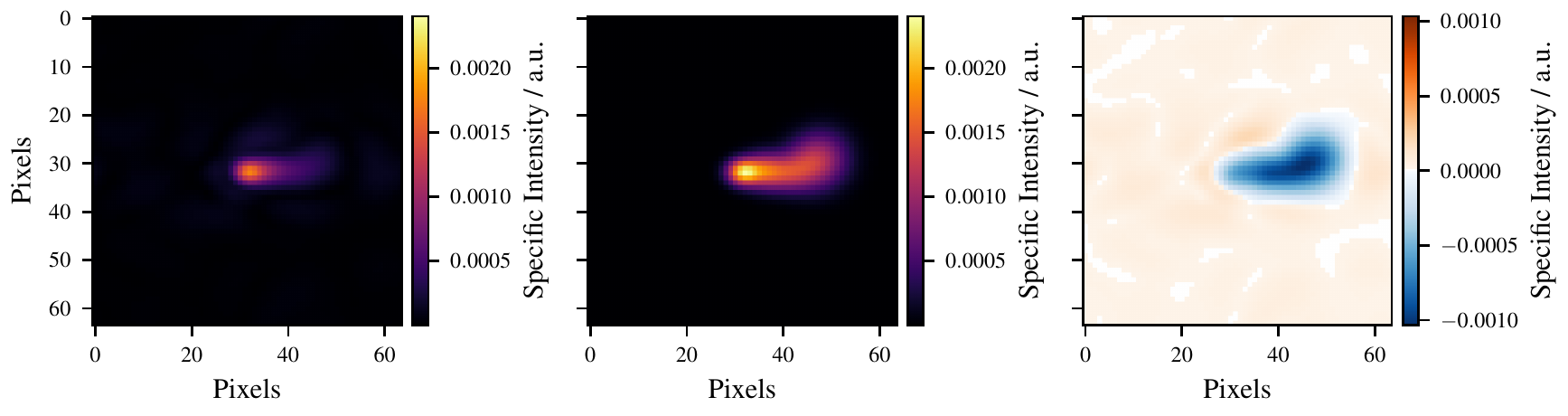}}
    \caption{Reconstruction of a one-sided jet with kinks and overlapping Gaussian components, which was not used in the training procedure. Resulting clean image (left), simulated brightness distribution (middle) and difference between both (right).}
    \label{fig:new_form_wo_fine_tune_one}
\end{figure*}

\begin{figure*}
    \centering
    \resizebox{\hsize}{!}
    {\includegraphics[width=\hsize]{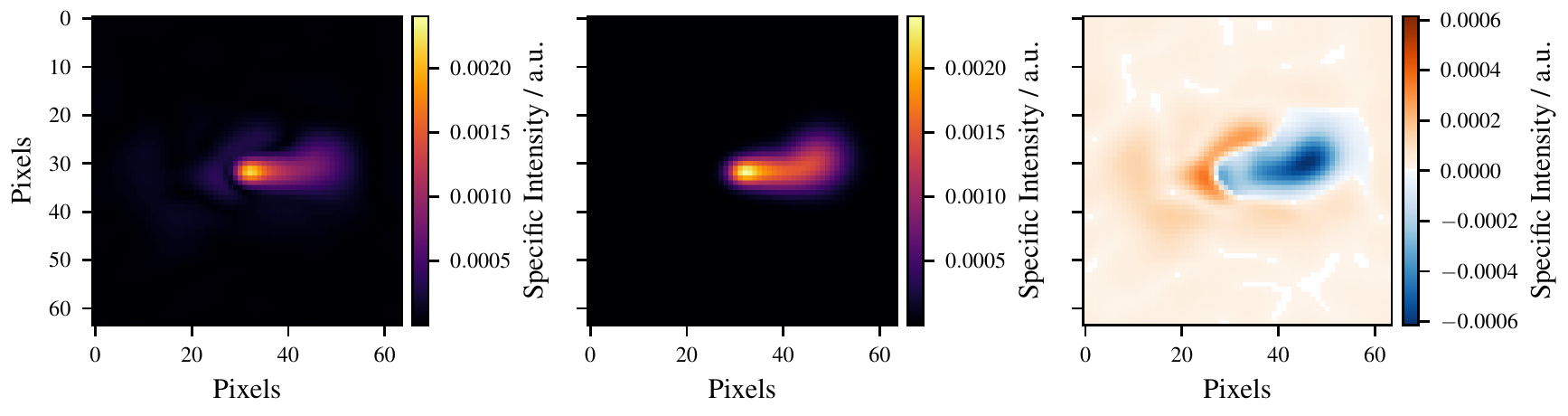}}
    \caption{Reconstruction of a one-sided jet with kinks and overlapping Gaussian components using the network after fine-tuning on the new source shapes. Resulting clean image (left), simulated brightness distribution (middle) and difference between both (right).}
    \label{fig:new_form_w_fine_tune_one}
\end{figure*}

\subsection{Additional Source Shapes}
\label{sec:new_shapes}

Machine learning algorithms learn from examples in the training data set and they use the acquired knowledge to make predictions for new examples. One of the limitations of machine learning is its inherent difficulty to generalize their acquired knowledge.

\begin{figure*}
    \centering
    \resizebox{\hsize}{!}
    {\includegraphics[width=\hsize]{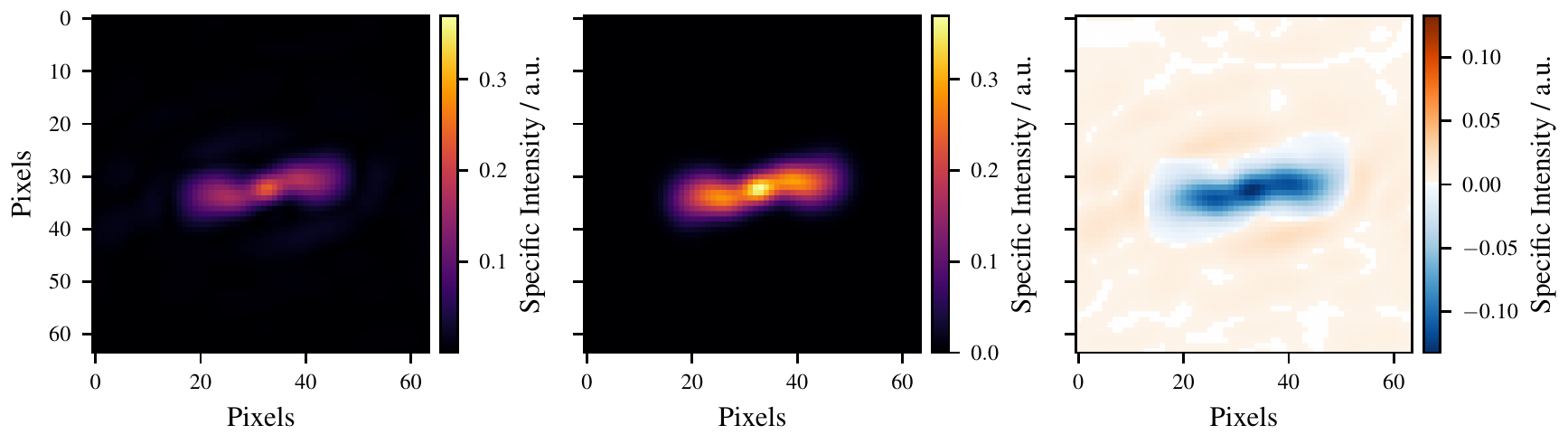}}
    \caption{Reconstruction of a two-sided jet with kinks and overlapping Gaussian components, which was not used in the training procedure. Resulting clean image (left), simulated brightness distribution (middle) and difference between both (right).}
    \label{fig:new_form_wo_fine_tune_two}
\end{figure*}

To test our network on untrained source shapes, we have used the network from \autoref{sec:results_noiseless} and applied it to more complex simulated radio galaxies. In these more complex examples, the jets of the sources are bent and individual components start to overlap as additional rotations of the jets are taken into account. \autoref{fig:new_form_wo_fine_tune_one} and \autoref{fig:new_form_wo_fine_tune_two} show the reconstruction of new jet shapes for a one- and a two-sided source, respectively. In both cases, the specific intensity is underestimated, while the general source shape is reconstructed well.

\begin{figure*}
    \centering
    \resizebox{\hsize}{!}
    {\includegraphics[width=\hsize]{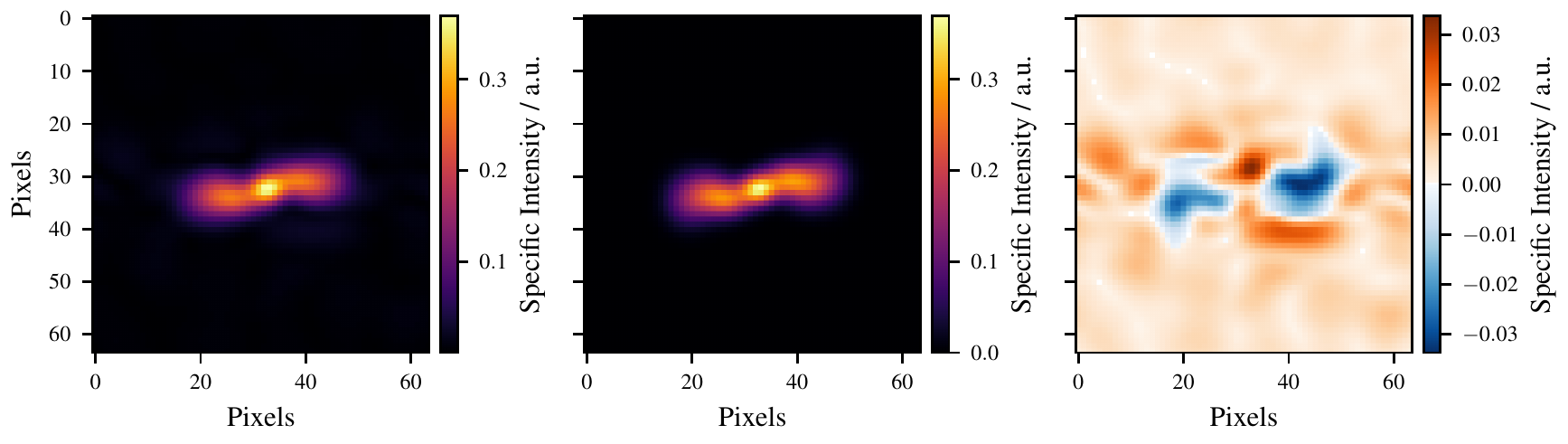}}
    \caption{Reconstruction of a two-sided jet with kinks and overlapping Gaussian components using the network after fine-tuning on the new source shapes. Resulting clean image (left), simulated brightness distribution (middle) and difference between both (right).}
    \label{fig:new_form_w_fine_tune_two}
\end{figure*}

Again, a fine-tuning of the existing Deep Learning network helps to increase the reconstruction quality. For this reason, we train the network for additional 40 epochs on a data set consisting of the new source shapes. This data set contains $10\,000$ training images and $2\,000$ validation images. The sampled amplitude and phase maps serve as input for our network. The fine tuning takes around $\SI{40}{\minute}$. \autoref{fig:new_form_w_fine_tune_one} and \autoref{fig:new_form_w_fine_tune_two} visualize the reconstruction of new jet shapes for the one- and the two-sided source reconstructed by the fine-tuned network, respectively. The reconstruction quality increased significantly as the intensity deviation between reconstruction and simulation decreased for both cases. Especially, fainter parts of the source are reconstructed better. This comes at the cost of more background artifacts.

For our network, this means that source shapes which were not used in the training process are less likely to be well-reconstructed. One advantage of radio interferometry is that many sources appear as two-dimensional Gaussian distributions or can be composed of several two-dimensional Gaussian distributions. Since the networks we have trained are capable of reconstructing Gaussian sources, they can do so to some extent on Gaussian sources with new shapes. However, when the jet components become very diffuse, the reconstruction quality drops significantly.

\section{Perspectives}
\label{sec:perspectives}

\subsection{Improved Data Preparation: Noise Estimations}

Analysis methods based on Deep Learning often have difficulties with very noisy input data. This problem mainly occurs because of the difficulty to describe the effect of noise on observations correctly in the simulations. This can be done by improving noise simulations or by improving the data preparation before the data enters the neural network.

Neural networks built from residual blocks show a reasonable capability for handling Gaussian noise. Nevertheless, further development and testing is necessary since noise on radio interferometer data leads to large uncertainties in the reconstructions and is a known problem in the conventional analysis.
To address this problem, we suggest a noise estimation and correction step already before the reconstruction step.
A possible solution is the application of another Deep Learning network in the data preparation step.

\subsection{Uncertainty Estimate}
\label{subsec:uncertainty}

A disadvantage of conventional imaging strategies is their inability to quantify the uncertainties in the reconstructed clean images.
To enable the Deep Learning network to quantify errors, we suggest an adjustment of the loss function, comparable to the one proposed in \cite{likelihood_loss}.
We assume a value originating from a Gaussian distribution with the parameters $\mu$ and $\sigma$ for every pixel. 
The negative log-likelihood of this Gaussian distribution, is 
\begin{align}
    -\mathcal{L} = 2 \log(\sigma) +  \frac{(x - \mu)^2}{\sigma^2},
\end{align}
and serves as a minimization function for the neural network.
Now, the network predicts two values for every pixel, where $\mu$ is the reconstructed value for this pixel, and $\sigma$ corresponds to the estimated uncertainty for this value.
$x$ represents the true value for the specific pixel. 
This approach has its limits, as the pixels are unavoidably correlated.
This loss function is an extension of the mean squared error (MSE), which results in the special case of $\sigma$ being the same for all pixels.

Estimates for $\mu$ and $\sigma$ make it possible to vary the reconstruction obtained by the neural network. The one-dimensional Gaussian distribution obtained for each pixel permits the sampling of several values for a specific pixel.
As a result, we obtain $n$ versions of the reconstructed amplitudes and phases. 
The pixel-wise comparison of these reconstructions helps to quantify uncertain regions by calculating the standard deviation of the $n$ different versions.
Furthermore, these $n$ versions of the reconstructions allow for creating $n$ reconstructed clean images which enable the quantification of uncertainties in image space. 

The first tests show how regions with high uncertainties in the resulting clean images can be identified.
In the future, we plan to extend this approach such that the quality of the reconstructions can be estimated even if the real brightness distribution is unknown.

\subsection{Source Finder and Source List Prediction}

In radio astronomy one may not always want to obtain an image but rather extract parameters such as the source position and the flux density.
Hence, alternative approaches may be more suitable for the analysis of images of large areas of the sky.
The cleaning process of such images puts high demands on computer memory.
We suggest the use of Deep Learning networks to estimate source parameters directly without obtaining an image first.
Architectures developed for object detection, e.g. the SSD300 architecture \citep{ssd300}, can help to solve this task.
These networks are able to infer source positions and, simultaneously, can classify the sources.
A loss function using the object's bounding boxes enables the estimation of an arbitrary number of sources in the input image \citep{bounding_boxes}. 
%The source lists obtained in this way are suitable for scientific analysis of the chosen sky region.

Finally, this approach allows for the improvement of conventional imaging software. During the reconstruction process, flux from source regions gets extracted iteratively in the form of point sources. 
In this way, common imagers create a model consisting of multiple point sources for the  source. Replacing the existing routines with an approach based on a Deep Learning network can help to improve the speed and the accuracy of the currently used methods.
\section{Conclusions}
\label{sec:conclusions}

A new generation of radio interferometers with enormous data rates requires analysis strategies that generate reproducible results on short timescales and with affordable computing resources.
A way to achieve this goal is to apply Deep Learning-based analysis methods directly on the \uv data.

Our simulation is designed to quickly generate images based on the appearance of cleaned radio interferometry measurements.
The \uv coverage of the simulated sources are sampled by \uv masks based on observations with the VLBA.
Thus, we generate incomplete \uv data as in realistic radio interferometric observations. This data serves as input for the neural network.

We have shown that our approach of using neural networks to reconstruct incomplete \uv spaces can generate reproducible results quickly and reliably in the case of radio galaxies made from Gaussian components. Deviations between reconstructed and simulated images vary depending on the quantity, we are looking at. Especially the jet angles and the source area ratios, as well as the mean specific intensity deviations of the core components are constructed fairly reliably. The corresponding mean values are summarized in \autoref{tab:results_noise_filling}. The results in the lower part of \autoref{tab:results_noise_filling} show that larger sampling densities in \uv space lead to better reconstructions by the neural network. Furthermore, our network can deal with noisy input data. This result is supported by the example images and histograms presented in \autoref{sec:noisy_input}. 

For comparison with an established cleaning method, we processed our data using \texttt{wsclean}. The results of \autoref{subsec:advanced-eval-methos} suggest that the reconstruction with \texttt{wsclean} does not perform as well as the Deep Learning networks. At this point we would like to emphasize that the direct comparison is not completely fair. The networks we have trained are designed to reconstruct a wide range of similar input data. This is made possible by the large statistics during the training process, which takes several hours, as described in \autoref{sec:training}.

When looking at derived quantities such as jet angles, it turns out that our neural network performs better than  \texttt{wsclean}.

The big advantage of \texttt{wsclean} is that there is no need to train a network first.
After finding suitable cleaning parameters, the creation of the clean image takes only a few seconds.
The disadvantage is that the quality of the reconstructed images depends strongly on the selected cleaning parameters and these differ for different input data. 
This makes it difficult to easily apply \texttt{wsclean} to a large data set even if the data quality of the input data hardly differs, see \autoref{subsec:advanced-eval-methos}.
Furthermore, finding suitable cleaning parameters can take many iterative approaches.

Another difference between the two methods is that our Deep Learning reconstructions are not convolved with a clean beam.
The input data is reconstructed directly in Fourier space and the clean image is created by the inverse Fourier transformer of the fully filled \uv space.
In case of \texttt{wsclean} a point source model is created which is then convolved with the theoretical clean beam.
This results in smearing of the reconstructed brightness distributions. This is partly due to technical obstacles, such as small image sizes, and partly due to the fact that our chosen parameters from \texttt{wsclean} are only optimized for a handful of images and therefore do not perform as well as our network on a \num{10000} image data set. Hence, our approach may have advantages over \texttt{wsclean} when it comes to fast reconstructing a large number of images.
%For this use case, the fast application times, listed in \autoref{tab:run-times} for different image sizes, are also of great advantage.

 In \autoref{sec:new_shapes}, we discuss the performance of a our network applied to a more complex data set with kinked and diffuse jet sources. The results show that our network is able to reconstruct these examples reasonably well.

In conclusion, we have made a proof-of-concept that Deep Learning methods can be applied to reconstruct incomplete Fourier data in radio interferometric imaging. Even input data with uncorrelated noise produces results that match those created with noiseless input data.

Our analysis framework \texttt{radionets} \citep{radionets} is made available as an open-source package. In future releases, we will improve our simulations by utilizing  RIME and the Jones calculus. 
Thus, we enable the consideration of additional complications such as the point spread function of the radio telescopes, the influence of side-lobes, or multi-channel data.
In combination with the ability to handle larger image sizes, this will open the way to train Deep Learning networks applicable to real data.
Furthermore, neural networks can be used to quantify uncertainties in the reconstructed clean images.  Uncertainty maps will help to evaluate the reconstruction results. More advanced segmentation techniques enhance the location of source positions. The segmentation maps can be used to improve the performance of established imaging software.

\begin{acknowledgements}
This work is supported by Deutsche Forschungsgemeinschaft (DFG) --project number 124020371-- within the Collaborative Research Center SFB 876 "Providing Information by Resource-Constrained Analysis", DFG project number 124020371, SFB project C3.
We acknowledge support from the BMBF (Verbundforschung).
MB acknowledges support from the Deutsche Forschungsgemeinschaft under Germany's Excellence Strategy - EXC 2121 "Quantum Universe" - 390833306, and the support and collaboration with the Center for Data and Computing in Natural Sciences (CDCS).
We thank Kai Brügge for his initial motivation and discussions that formed the foundation for this work.

We thank Richard Wiemann for fruitful discussions and helpful ideas.
\end{acknowledgements}

\bibliographystyle{aa}
\bibliography{references}

\begin{thebibliography}{51}
\expandafter\ifx\csname natexlab\endcsname\relax\def\natexlab#1{#1}\fi

\bibitem[{Abbasi {et~al.}(2021)Abbasi, Ackermann, Adams, Aguilar, Ahlers,
  Ahrens, Alispach, au2, Amin, An, Andeen, Anderson, Ansseau, Anton,
  Argüelles, Axani, Bai, V., Barbano, Barwick, Bastian, Basu, Baum, Baur, Bay,
  Beatty, Becker, Tjus, Bellenghi, BenZvi, Berley, Bernardini, Besson, Binder,
  Bindig, Blaufuss, Blot, Böser, Botner, Böttcher, Bourbeau, Bourbeau,
  Bradascio, Braun, Bron, Brostean-Kaiser, Burgman, Busse, Campana, Chen,
  Chirkin, Choi, Clark, Clark, Classen, Coleman, Collin, Conrad, Coppin,
  Correa, Cowen, Cross, Dave, Clercq, DeLaunay, Dembinski, Deoskar, Ridder,
  Desai, Desiati, de~Vries, de~Wasseige, de~With, DeYoung, Dharani, Diaz,
  Díaz-Vélez, Dujmovic, Dunkman, DuVernois, Dvorak, Ehrhardt, Eller, Engel,
  Evans, Evenson, Fahey, Fazely, Fiedlschuster, Fienberg, Filimonov, Finley,
  Fischer, Fox, Franckowiak, Friedman, Fritz, Fürst, Gaisser, Gallagher,
  Ganster, Garrappa, Gerhardt, Ghadimi, Glaser, Glauch, Glüsenkamp,
  Goldschmidt, Gonzalez, Goswami, Grant, Grégoire, Griffith, Griswold,
  Gündüz, Haack, Hallgren, Halliday, Halve, Halzen, Minh, Hanson, Hardin,
  Harnisch, Haungs, Hauser, Hebecker, Helbing, Henningsen, Hettinger, Hickford,
  Hignight, Hill, Hill, Hoffman, Hoffmann, Hoinka, Hokanson-Fasig, Hoshina,
  Huang, Huber, Huber, Hultqvist, Hünnefeld, Hussain, In, Iovine, Ishihara,
  Jansson, Japaridze, Jeong, Jones, Joppe, Kang, Kang, Kang, Kappes, Kappesser,
  Karg, Karl, Karle, Katz, Kauer, Kellermann, Kelley, Kheirandish, Kim, Kin,
  Kintscher, Kiryluk, Klein, Koirala, Kolanoski, Köpke, Kopper, Kopper,
  Koskinen, Koundal, Kovacevich, Kowalski, Krings, Krückl, Kurahashi,
  Kyriacou, Gualda, Lanfranchi, Larson, Lauber, Lazar, Leonard, Leszczyńska,
  Li, Liu, Lohfink, Mariscal, Lu, Lucarelli, Ludwig, Luszczak, Lyu, Ma, Madsen,
  Mahn, Makino, Mallik, Mancina, Mari{ş}, Maruyama, Mase, McNally, Meagher,
  Medina, Meier, Meighen-Berger, Merz, Micallef, Mockler, Momenté, Montaruli,
  Moore, Morik, Morse, Moulai, Naab, Nagai, Naumann, Necker, Nguy{\~{ê}}n,
  Niederhausen, Nisa, Nowicki, Nygren, Pollmann, Oehler, Olivas, O'Sullivan,
  Pandya, Pankova, Park, Parker, Paudel, Peiffer, de~los Heros, Philippen,
  Pieloth, Pieper, Pizzuto, Plum, Popovych, Porcelli, Rodriguez, Price, Pries,
  Przybylski, Raab, Raissi, Rameez, Rawlins, Rea, Rehman, Reimann, Renschler,
  Renzi, Resconi, Reusch, Rhode, Richman, Riedel, Robertson, Roellinghoff,
  Rongen, Rott, Ruhe, Ryckbosch, Cantu, Safa, Herrera, Sandrock, Sandroos,
  Santander, Sarkar, Sarkar, Satalecka, Scharf, Schaufel, Schieler, Schlunder,
  Schmidt, Schneider, Schneider, Schröder, Schumacher, Sclafani, Seckel,
  Seunarine, Sharma, Shefali, Silva, Skrzypek, Smithers, Snihur, Soedingrekso,
  Soldin, Spiczak, Spiering, Stachurska, Stamatikos, Stanev, Stein, Stettner,
  Steuer, Stezelberger, Stokstad, Stürwald, Stuttard, Sullivan, Taboada,
  Tenholt, Ter-Antonyan, Tilav, Tischbein, Tollefson, Tomankova, Tönnis,
  Toscano, Tosi, Trettin, Tselengidou, Tung, Turcati, Turcotte, Turley,
  Twagirayezu, Ty, Elorrieta, Valtonen-Mattila, Vandenbroucke, van Eijk, van
  Eijndhoven, Vannerom, van Santen, Verpoest, Vraeghe, Walck, Wallace, Watson,
  Weaver, Weindl, Weiss, Weldert, Wendt, Werthebach, Weyrauch, Whelan,
  Whitehorn, Wiebe, Wiebusch, Williams, Wolf, Woschnagg, Wrede, Wulff, Xu, Xu,
  Yanez, Yoshida, Yuan, \& Zhang}]{likelihood_loss}
Abbasi, R., Ackermann, M., Adams, J., {et~al.} 2021, A Convolutional Neural
  Network based Cascade Reconstruction for the IceCube Neutrino Observatory

\bibitem[{Amari(1993)}]{sgd}
Amari, S.-I. 1993, Neurocomputing, 5, 185

\bibitem[{{Blandford} \& {K{\"o}nigl}(1979)}]{bland_koenigl}
{Blandford}, R.~D. \& {K{\"o}nigl}, A. 1979, \apj, 232, 34

\bibitem[{Bridle \& Cohen(2012)}]{radiojets}
Bridle, A.~H. \& Cohen, M.~H. 2012, Observational Details: Radio (John Wiley \&
  Sons, Ltd), 115--152

\bibitem[{Broten {et~al.}(1967)Broten, Legg, Locke, McLeish, Richards,
  Chisholm, Gush, Yen, \& Galt}]{vlbi_broten}
Broten, N.~W., Legg, T.~H., Locke, J.~L., {et~al.} 1967, Science, 156, 1592

\bibitem[{Clark(1980)}]{clean}
Clark, B.~G. 1980, \aap, 89, 377

\bibitem[{{Erhan} {et~al.}(2013){Erhan}, {Szegedy}, {Toshev}, \&
  {Anguelov}}]{bounding_boxes}
{Erhan}, D., {Szegedy}, C., {Toshev}, A., \& {Anguelov}, D. 2013, arXiv
  e-prints, arXiv:1312.2249

\bibitem[{Ghirlanda {et~al.}(2019)Ghirlanda, Salafia, Paragi, Giroletti, Yang,
  Marcote, Blanchard, Agudo, An, Bernardini, Beswick, Branchesi, Campana,
  Casadio, Chassande-Mottin, Colpi, Covino, D{\textquoteright}Avanzo,
  D{\textquoteright}Elia, Frey, Gawronski, Ghisellini, Gurvits, Jonker, van
  Langevelde, Melandri, Moldon, Nava, Perego, Perez-Torres, Reynolds,
  Salvaterra, Tagliaferri, Venturi, Vergani, \& Zhang}]{mwl_gw}
Ghirlanda, G., Salafia, O.~S., Paragi, Z., {et~al.} 2019, Science, 363, 968

\bibitem[{{Grainge} {et~al.}(2017){Grainge}, {Alachkar}, {Amy}, {Barbosa},
  {Bommineni}, {Boven}, {Braddock}, {Davis}, {Diwakar}, {Francis},
  {Gabrielczyk}, {Gamatham}, {Garrington}, {Gibbon}, {Gozzard}, {Gregory},
  {Guo}, {Gupta}, {Hammond}, {Hindley}, {Horn}, {Hughes-Jones}, {Hussey},
  {Lloyd}, {Mammen}, {Miteff}, {Mohile}, {Muller}, {Natarajan}, {Nicholls},
  {Oberland}, {Pearson}, {Rayner}, {Schediwy}, {Schilizzi}, {Sharma}, {Stobie},
  {Tearle}, {Wang}, {Wallace}, {Wang}, {Warange}, {Whitaker}, {Wilkinson}, \&
  {Wingfield}}]{ska}
{Grainge}, K., {Alachkar}, B., {Amy}, S., {et~al.} 2017, Astronomy Reports, 61,
  288

\bibitem[{Gross \& Wilber(2016)}]{residual-block}
Gross, S. \& Wilber, M. 2016, Training and investigating Residual Nets

\bibitem[{Hastie {et~al.}(2009)Hastie, Tibshirani, \& Friedman}]{hastie}
Hastie, T., Tibshirani, R., \& Friedman, J. 2009, The elements of statistical
  learning: data mining, inference and prediction, 2nd edn. (Springer)

\bibitem[{He {et~al.}(2015{\natexlab{a}})He, Zhang, Ren, \&
  Sun}]{residual-nets}
He, K., Zhang, X., Ren, S., \& Sun, J. 2015{\natexlab{a}}, Deep Residual
  Learning for Image Recognition

\bibitem[{He {et~al.}(2015{\natexlab{b}})He, Zhang, Ren, \& Sun}]{prelu}
He, K., Zhang, X., Ren, S., \& Sun, J. 2015{\natexlab{b}}, in Proceedings of
  the IEEE International Conference on Computer Vision (ICCV)

\bibitem[{Howard \& Gugger(2020)}]{lrfind}
Howard, J. \& Gugger, S. 2020, Deep Learning for Coders with fastai and PyTorch
  (O'Reilly Media, Inc.), 205--207

\bibitem[{Howard {et~al.}(2018)}]{fastai}
Howard, J. {et~al.} 2018, fastai, \url{https://github.com/fastai/fastai}

\bibitem[{{Hunter}(2007)}]{matplotlib}
{Hunter}, J.~D. 2007, Computing in Science Engineering, 9, 90

\bibitem[{{Jaeger}(2008)}]{casa}
{Jaeger}, S. 2008, in Astronomical Society of the Pacific Conference Series,
  Vol. 394, Astronomical Data Analysis Software and Systems XVII, ed. R.~W.
  {Argyle}, P.~S. {Bunclark}, \& J.~R. {Lewis}, 623

\bibitem[{Kingma \& Ba(2017)}]{adam}
Kingma, D.~P. \& Ba, J. 2017, Adam: A Method for Stochastic Optimization

\bibitem[{Königl(1980)}]{koenigl}
Königl, A. 1980, The Physics of Fluids, 23, 1083

\bibitem[{Ledig {et~al.}(2016)Ledig, Theis, Huszar, Caballero, Cunningham,
  Acosta, Aitken, Tejani, Totz, Wang, \& Shi}]{superres}
Ledig, C., Theis, L., Huszar, F., {et~al.} 2016, Photo-Realistic Single Image
  Super-Resolution Using a Generative Adversarial Network

\bibitem[{Linhoff {et~al.}(2020)Linhoff, Sandrock, Kadler, Elsässer, \&
  Rhode}]{mwl_lena}
Linhoff, L., Sandrock, A., Kadler, M., Elsässer, D., \& Rhode, W. 2020,
  Monthly Notices of the Royal Astronomical Society, 500, 4671

\bibitem[{Lister {et~al.}(2013)Lister, Aller, Aller, Homan, Kellermann,
  Kovalev, Pushkarev, Richards, Ros, \& Savolainen}]{gaussian_1}
Lister, M.~L., Aller, M.~F., Aller, H.~D., {et~al.} 2013, The Astronomical
  Journal, 146, 120

\bibitem[{Lister {et~al.}(2019)Lister, Homan, Hovatta, Kellermann, Kiehlmann,
  Kovalev, Max-Moerbeck, Pushkarev, Readhead, Ros, \& Savolainen}]{gaussian_2}
Lister, M.~L., Homan, D.~C., Hovatta, T., {et~al.} 2019, The Astrophysical
  Journal, 874, 43

\bibitem[{Liu {et~al.}(2016)Liu, Anguelov, Erhan, Szegedy, Reed, Fu, \&
  Berg}]{ssd300}
Liu, W., Anguelov, D., Erhan, D., {et~al.} 2016, in Computer Vision -- ECCV
  2016, ed. B.~Leibe, J.~Matas, N.~Sebe, \& M.~Welling (Cham: Springer
  International Publishing), 21--37

\bibitem[{McKinney {et~al.}(2010)}]{pandas}
McKinney, W. {et~al.} 2010, in Proceedings of the 9th Python in Science
  Conference, Vol. 445, Austin, TX, 51--56

\bibitem[{Morningstar {et~al.}(2018)Morningstar, Hezaveh, Perreault~Levasseur,
  Blandford, Marshall, Putzky, \& Wechsler}]{rim1}
Morningstar, W.~R., Hezaveh, Y.~D., Perreault~Levasseur, L., {et~al.} 2018,
  arXiv e-prints [\eprint[arXiv]{1808.00011}]

\bibitem[{Morningstar {et~al.}(2019)Morningstar, Perreault~Levasseur, Hezaveh,
  Blandford, Marshall, Putzky, Rueter, Wechsler, \& Welling}]{rim2}
Morningstar, W.~R., Perreault~Levasseur, L., Hezaveh, Y.~D., {et~al.} 2019,
  arXiv e-prints [\eprint[arXiv]{1901.01359}]

\bibitem[{{Napier} {et~al.}(1994){Napier}, {Bagri}, {Clark}, {Rogers},
  {Romney}, {Thompson}, \& {Walker}}]{vlba}
{Napier}, P.~J., {Bagri}, D.~S., {Clark}, B.~G., {et~al.} 1994, Proceedings of
  the IEEE, 82, 658

\bibitem[{Needham(1999)}]{euler}
Needham, T. 1999, Visual Complex Analysis (Oxford University Press), 10--14

\bibitem[{Office(2011-2018)}]{cartopy}
Office, M. 2011-2018, Cartopy

\bibitem[{Offringa {et~al.}(2014)Offringa, McKinley, Hurley-Walker, Briggs,
  Wayth, Kaplan, Bell, Feng, Neben, Hughes, Rhee, Murphy, Bhat, Bernardi,
  Bowman, Cappallo, Corey, Deshpande, Emrich, Ewall-Wice, Gaensler, Goeke,
  Greenhill, Hazelton, Hindson, Johnston-Hollitt, Jacobs, Kasper, Kratzenberg,
  Lenc, Lonsdale, Lynch, McWhirter, Mitchell, Morales, Morgan, Kudryavtseva,
  Oberoi, Ord, Pindor, Procopio, Prabu, Riding, Roshi, Shankar, Srivani,
  Subrahmanyan, Tingay, Waterson, Webster, Whitney, Williams, \&
  Williams}]{wsclean}
Offringa, A.~R., McKinley, B., Hurley-Walker, N., {et~al.} 2014, Monthly
  Notices of the Royal Astronomical Society, 444, 606

\bibitem[{Oliphant(2006)}]{numpy}
Oliphant, T.~E. 2006, A guide to NumPy, Vol.~1 (Trelgol Publishing USA)

\bibitem[{Ostrovsky {et~al.}(2009)Ostrovsky, Mart\'{i}nez-Niconoff,
  Mart\'{i}nez-Vara, \& Olvera-Santamar\'{i}a}]{Cittert-Zernike}
Ostrovsky, A.~S., Mart\'{i}nez-Niconoff, G., Mart\'{i}nez-Vara, P., \&
  Olvera-Santamar\'{i}a, M.~A. 2009, Opt. Express, 17, 1746

\bibitem[{Paszke {et~al.}(2019)Paszke, Gross, Massa, Lerer, Bradbury, Chanan,
  Killeen, Lin, Gimelshein, Antiga, Desmaison, Kopf, Yang, DeVito, Raison,
  Tejani, Chilamkurthy, Steiner, Fang, Bai, \& Chintala}]{pytorch}
Paszke, A., Gross, S., Massa, F., {et~al.} 2019, in Advances in Neural
  Information Processing Systems 32, ed. H.~Wallach, H.~Larochelle,
  A.~Beygelzimer, F.~d\textquotesingle Alch\'{e}-Buc, E.~Fox, \& R.~Garnett
  (Curran Associates, Inc.), 8026--8037

\bibitem[{Price-Whelan {et~al.}(2018)Price-Whelan, Sip{\H{o}}cz, Günther, Lim,
  Crawford, Conseil, Shupe, Craig, Dencheva, Ginsburg, VanderPlas, Bradley,
  P{\'{e}}rez-Su{\'{a}}rez, de~Val-Borro, Aldcroft, Cruz, Robitaille, Tollerud,
  Ardelean, Babej, Bach, Bachetti, Bakanov, Bamford, Barentsen, Barmby,
  Baumbach, Berry, Biscani, Boquien, Bostroem, Bouma, Brammer, Bray,
  Breytenbach, Buddelmeijer, Burke, Calderone, Rodr{\'{\i}}guez, Cara, Cardoso,
  Cheedella, Copin, Corrales, Crichton, D'Avella, Deil, Depagne, Dietrich,
  Donath, Droettboom, Earl, Erben, Fabbro, Ferreira, Finethy, Fox, Garrison,
  Gibbons, Goldstein, Gommers, Greco, Greenfield, Groener, Grollier, Hagen,
  Hirst, Homeier, Horton, Hosseinzadeh, Hu, Hunkeler, Ivezi{\'{c}}, Jain,
  Jenness, Kanarek, Kendrew, Kern, Kerzendorf, Khvalko, King, Kirkby, Kulkarni,
  Kumar, Lee, Lenz, Littlefair, Ma, Macleod, Mastropietro, McCully, Montagnac,
  Morris, Mueller, Mumford, Muna, Murphy, Nelson, Nguyen, Ninan, Nöthe, Ogaz,
  Oh, Parejko, Parley, Pascual, Patil, Patil, Plunkett, Prochaska, Rastogi,
  Janga, Sabater, Sakurikar, Seifert, Sherbert, Sherwood-Taylor, Shih, Sick,
  Silbiger, Singanamalla, Singer, Sladen, Sooley, Sornarajah, Streicher,
  Teuben, Thomas, Tremblay, Turner, Terr{\'{o}}n, van Kerkwijk, de~la Vega,
  Watkins, Weaver, Whitmore, Woillez, \& Zabalza}]{astropy2}
Price-Whelan, A.~M., Sip{\H{o}}cz, B.~M., Günther, H.~M., {et~al.} 2018, The
  Astronomical Journal, 156, 123

\bibitem[{Python{\ }Software{\ }Foundation(2020)}]{python}
Python{\ }Software{\ }Foundation. 2020, Python

\bibitem[{Renard {et~al.}(2010)Renard, Malbet, Benisty, Thi\'ebaut, \&
  Berger}]{milli-arc}
Renard, S., Malbet, F., Benisty, M., Thi\'ebaut, E., \& Berger, J.-P. 2010,
  A\&A, 519, A26

\bibitem[{Schmidt {et~al.}(2019)Schmidt, Geyer, {et~al.}}]{radionets}
Schmidt, K., Geyer, F., {et~al.} 2019, radionets,
  \url{https://github.com/radionets-project/radionets}

\bibitem[{Shepherd {et~al.}(1994)Shepherd, Pearson, \& Taylor}]{difmap}
Shepherd, M.~C., Pearson, T.~J., \& Taylor, G.~B. 1994, Bulletin of the
  Astronomical Society, 26, 987

\bibitem[{Smirnov(2011)}]{rime}
Smirnov, O.~M. 2011, A\&A, 527, A106

\bibitem[{Taylor {et~al.}(1999)Taylor, Carilli, \& Perley}]{rotation}
Taylor, G.~B., Carilli, C.~L., \& Perley, R.~A. 1999, in Astronomical Society
  of the Pacific Conference Series, Vol. 180, Synthesis Imaging in Radio
  Astronomy II

\bibitem[{{The Astropy Collaboration} {et~al.}(2013){The Astropy
  Collaboration}, {Robitaille, Thomas P.}, {Tollerud, Erik J.}, {Greenfield,
  Perry}, {Droettboom, Michael}, {Bray, Erik}, {Aldcroft, Tom}, {Davis, Matt},
  {Ginsburg, Adam}, {Price-Whelan, Adrian M.}, {Kerzendorf, Wolfgang E.},
  {Conley, Alexander}, {Crighton, Neil}, {Barbary, Kyle}, {Muna, Demitri},
  {Ferguson, Henry}, {Grollier, Fr\'ed\'eric}, {Parikh, Madhura M.}, {Nair,
  Prasanth H.}, {G\"unther, Hans M.}, {Deil, Christoph}, {Woillez, Julien},
  {Conseil, Simon}, {Kramer, Roban}, {Turner, James E. H.}, {Singer, Leo},
  {Fox, Ryan}, {Weaver, Benjamin A.}, {Zabalza, Victor}, {Edwards, Zachary I.},
  {Azalee Bostroem, K.}, {Burke, D. J.}, {Casey, Andrew R.}, {Crawford, Steven
  M.}, {Dencheva, Nadia}, {Ely, Justin}, {Jenness, Tim}, {Labrie, Kathleen},
  {Lim, Pey Lian}, {Pierfederici, Francesco}, {Pontzen, Andrew}, {Ptak, Andy},
  {Refsdal, Brian}, {Servillat, Mathieu}, \& {Streicher, Ole}}]{astropy1}
{The Astropy Collaboration}, {Robitaille, Thomas P.}, {Tollerud, Erik J.},
  {et~al.} 2013, A\&A, 558, A33

\bibitem[{Thompson {et~al.}(2001{\natexlab{a}})Thompson, Moran, \&
  Swenson~Jr.}]{ri_clean}
Thompson, A.~R., Moran, J.~M., \& Swenson~Jr., G.~W. 2001{\natexlab{a}},
  Introductory Theory of Interferometry and Synthesis Imaging (John Wiley \&
  Sons, Ltd), 72

\bibitem[{Thompson {et~al.}(2001{\natexlab{b}})Thompson, Moran, \&
  Swenson~Jr.}]{ri_intro}
Thompson, A.~R., Moran, J.~M., \& Swenson~Jr., G.~W. 2001{\natexlab{b}},
  Introductory Theory of Interferometry and Synthesis Imaging (John Wiley \&
  Sons, Ltd), 50--67

\bibitem[{Thompson {et~al.}(2001{\natexlab{c}})Thompson, Moran, \&
  Swenson~Jr.}]{ri_cittert}
Thompson, A.~R., Moran, J.~M., \& Swenson~Jr., G.~W. 2001{\natexlab{c}}, Van
  Cittert‐Zernike Theorem, Spatial Coherence, and Scattering (John Wiley \&
  Sons, Ltd), 594--612

\bibitem[{Urry \& Padovani(1995)}]{urry}
Urry, C. \& Padovani, P. 1995, Publications of the Astronomical Society of the
  Pacific, 107, 803

\bibitem[{Van~der Walt {et~al.}(2014)Van~der Walt, Sch{\"o}nberger,
  Nunez-Iglesias, Boulogne, Warner, Yager, Gouillart, \& Yu}]{scikit-image}
Van~der Walt, S., Sch{\"o}nberger, J.~L., Nunez-Iglesias, J., {et~al.} 2014,
  PeerJ, 2, e453

\bibitem[{van~der Walt {et~al.}(2014)van~der Walt, {S}ch\"onberger,
  {Nunez-Iglesias}, {B}oulogne, {W}arner, {Y}ager, {G}ouillart, {Y}u, \& the
  scikit-image contributors}]{skimage}
van~der Walt, S., {S}ch\"onberger, J.~L., {Nunez-Iglesias}, J., {et~al.} 2014,
  PeerJ, 2, e453

\bibitem[{{van Haarlem} {et~al.}(2013)}]{lofar}
{van Haarlem} {et~al.} 2013, A\&A, 556, A2

\bibitem[{Walter(1990)}]{leibniz}
Walter, W. 1990, Analysis I, 2. Auflage (Springer-Verlag Berling Heidelberg
  GmbH), 285 -- 286

\bibitem[{{Yang} {et~al.}(2019){Yang}, {Zhang}, {Tian}, {Wang}, {Xue}, \&
  {Liao}}]{superr}
{Yang}, W., {Zhang}, X., {Tian}, Y., {et~al.} 2019, IEEE Transactions on
  Multimedia, 21, 3106

\end{thebibliography}

\FloatBarrier
\begin{appendix}

\section{Used Software and Packages}

In this work, we use \texttt{PyTorch} \citep{pytorch} as the fundamental Deep Learning framework. It was choosen because of its flexibility and the ability to directly develop algorithms in the programming language \texttt{Python} \citep{python}. On top of \texttt{PyTorch} we use the Deep Learning library \texttt{fast.ai} \citep{fastai} which supplies high-level components to build customized Deep Learning algorithms in a quick and efficient way.
For simulations and data analysis the Python packages \texttt{NumPy} \citep{numpy}, \texttt{Astropy} \citep{astropy1, astropy2}, \texttt{Cartopy} \citep{cartopy}, \texttt{scikit-image} \citep{scikit-image} and \texttt{Pandas} \citep{pandas} are used.
The illustration of the results was done using the plotting package \texttt{Matplotlib} \citep{matplotlib}.

A full list of the used packages and our developed open-source \texttt{radionets} framework can be found on github: https://github.com/radionets-project/radionets

\section{Flux Distributions}

\begin{figure}[!ht]
\centering
\begin{minipage}{\textwidth}
    \centering
    \includegraphics[width=\textwidth,clip]{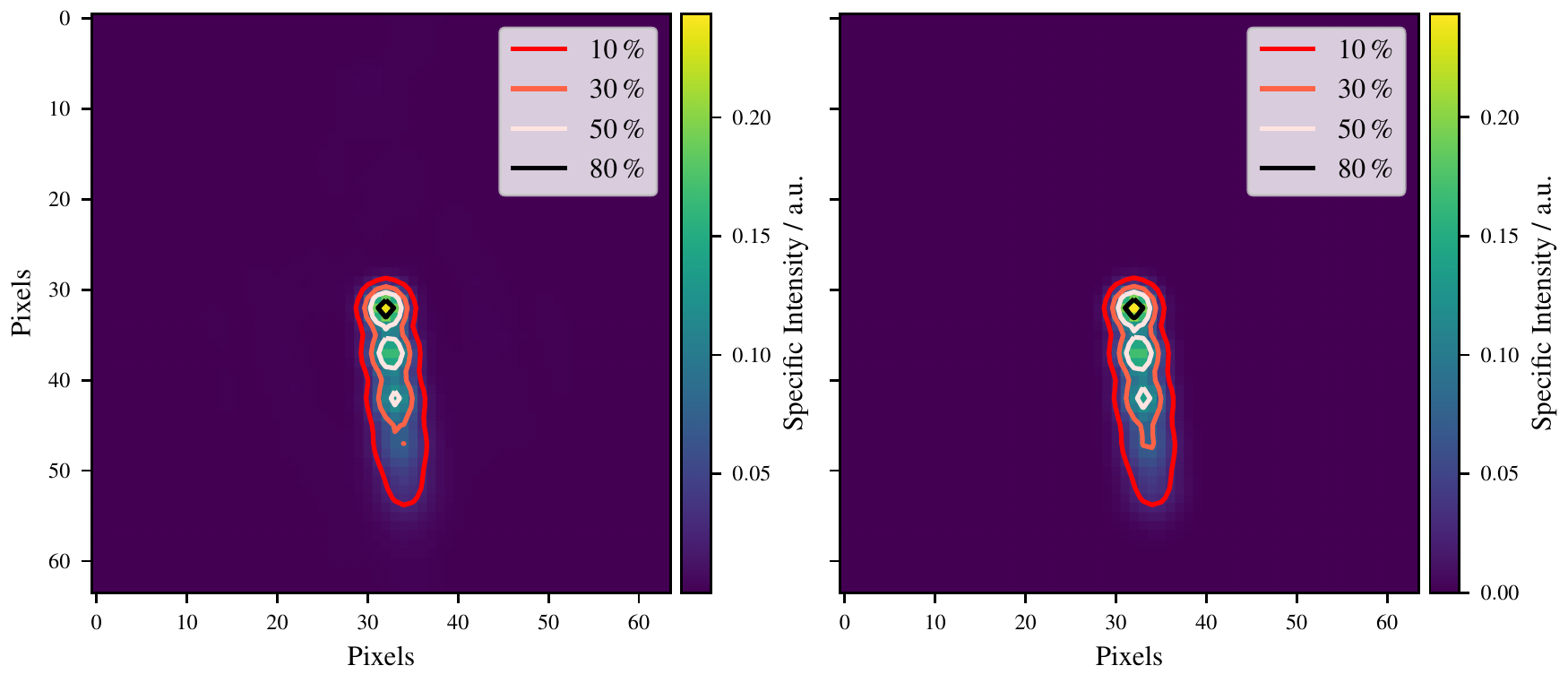}
    \caption{Exemplary contour plot for a training session with clean input images. Resulting clean image (left) and simulated brightness distribution (right) are shown. For both images, the contour levels are based on the peak flux density of the simulated brightness distribution. The ratio of 0.98 is calculated between the 10\% boundary of the prediction and the truth.}
    \label{fig:contour-plot-clean}
\end{minipage}
\end{figure}

\begin{figure*}[!ht]
\centering
\begin{minipage}{\textwidth}
    \centering
    \includegraphics[width=\textwidth,clip]{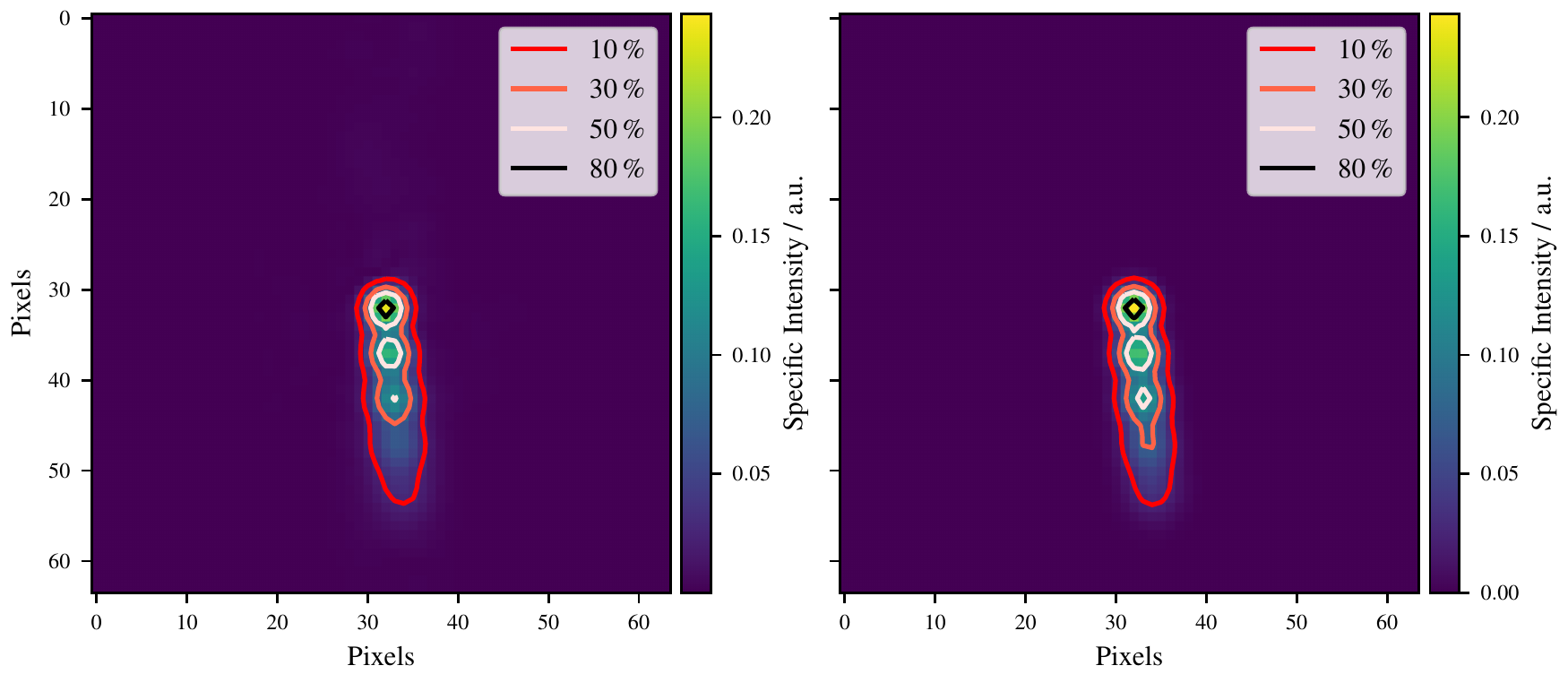}
    \caption{Exemplary contour plot for a training session with noisy input images. Resulting clean image (left) and simulated brightness distribution (right) are shown. For both images, the contour levels are based on the peak flux density of the simulated brightness distribution. The ratio of 1.12 is calculated between the 10\% boundary of the prediction and the truth.}
    \label{fig:contour-plot-noise}
\end{minipage}
\end{figure*}

\FloatBarrier

\section{Computer Setup}

\begin{table}[ht!]
    \centering
    \caption{Computer specifications for the setup used in the training process}
    \begin{tabular}{c|c|c}
    \toprule
    Part & Specification & Value \\
    \midrule
    GPU &  Nvidia GeForce RTX 2080 & \SI{8}{\gigabyte}\\
    CPU & Intel Core i7-8700k & 12 Cores @ \SI{3.7}{\giga\hertz}\\
    Hard drive & SSD & \SI{512}{\gigabyte} \\
    \bottomrule
    \end{tabular}
    \label{tab:computer-specifications}
\end{table}

\end{appendix}

%%%%%%%%%%%%%%%%%%%%%%%%%%%%%%%%%%%%%%%%%%%%%%%%%%

\end{document}